\documentclass[10pt]{book}
\usepackage{pgstylestandalone}

\title{Political Geometry}
\author{  }
\date{ }

\begin{document}
\Urlmuskip=0mu plus 1mu\relax % To make URLs linebreak

%\maketitle

 \chapter[Political Geography and Representation -- Rodden and Weighill]{Political Geography and Representation: A Case Study of Districting in Pennsylvania}
\label{RoddenWeighill}
\chapterauthor{JONATHAN RODDEN \\ THOMAS WEIGHILL}
\authorheader{RODDEN \& WEIGHILL}
\titleheader{Political Geography and Representation}

%\startcontents[chapters]
%\printcontents[chapters]{}{1}{}

\chaptersummary{This preprint offers a detailed look, both qualitative and quantitative, at districting with respect to recent voting patterns in one state: Pennsylvania. We investigate how much the partisan playing field is tilted by political geography. In particular we closely examine the role of scale. We find that partisan-neutral maps rarely give seats proportional to votes, and that making the district size smaller tends to make it even harder to find a proportional map. This preprint was prepared as a chapter in the forthcoming edited volume Political Geometry, an interdisciplinary collection of essays on redistricting.  (\url{mggg.org/gerrybook})}

%\thanks{Professor, Department of Political Science; Senior Fellow, Hoover Institution; and Senior Fellow, SIEPR, Stanford University {\tt jrodden@stanford.edu}}, Thomas Weighill\thanks{MGGG postdoc, Tufts University {\tt thomas.weighill@tufts.edu}}}

\section{Introduction}
The impressive success of recent ballot initiatives in Michigan, Missouri, Colorado, Ohio and Utah demonstrate that redistricting reform is broadly popular with voters.  But there is little agreement about what type of reform is optimal, and little knowledge about what is at stake when choosing between alternatives.  One type of reform, referred to by political and legal theorists as {\em process-oriented} reform, focuses on creating a redistricting process that is fair and transparent, giving authority to either independent commissioners or equal numbers of Democrats and Republicans, and encouraging them to pay little attention to the potential political outcomes associated with alternative maps, or perhaps even forbidding the analysis of electoral data altogether.  An alternative approach to reform is to focus explicitly on partisan outcomes---encouraging or requiring commissioners to draw maps that are fair to both parties according to some agreed criteria about how votes should translate into seats.    

Advocates of an outcome-based approach point out that while a pure process-oriented approach may be easy to explain and implement, it does not necessarily satisfy all the definitions of fairness stakeholders may have in mind. For example, many citizen observers prefer outcomes where the seat share for the parties matches the vote share (so-called ``proportional" outcomes), but neutral redistricting tends to lead to maps that result in representation that is far from proportional.  This phenomenon is the result of the particular way votes are distributed within a state (what we will refer to as \emph{political geography} in this chapter). This distribution has a signature form in the United States: Democrats are often highly concentrated in city centers and educated suburbs and Republicans are more dispersed in exurbs and rural areas \cite{rodden2019}.  

The partisan tendencies of neutral redistricting, and hence the stakes of debates about redistricting reform, are driven by each state's political geography, and perhaps especially by its urban political geography. An influential paper calling attention to quantifying the partisan tendencies of neutral redistricting was \cite{chenrodden2013}, which dubbed the phenomenon ``unintentional gerrymandering.'' The size, structure, and geographic arrangement of cities is extremely important for political representation in the United States \cite{rodden2019}. Moreover, it has been argued that the impact of urban geography on representation is conditioned by the geographic scale at which districts are drawn, as well as the overall level of support for the two parties \cite{eubankrodden2018}.  

Adding apparently neutral criteria like ``competitiveness'' \cite{defordduchinsolomon2019} or the procedure for defining adjacency across bodies of water \cite{mgggalaska2019} can have unforeseen and sometimes dramatic effects on the partisan statistics of maps drawn under a neutral process. See also \cite{mgggvacriteria} for an extensive survey of the impact of redistricting criteria for Virginia, including compactness, population balance, racial balance and locality splits. Also included in that paper are redistricting criteria which explicitly depend on vote data such as efficiency gap and mean-median scores. It is very clear that a neutral redistricting process that focuses only on creating compact, contiguous districts and minimizing county or municipal splits, for instance, can lead to what many would deem to be a ``fair'' outcome in some states but not in others, particularly if the definition of ``fair'' under consideration is based on the ability to translate overall vote share into seat share. Moreover, the fairness or lack thereof can depend in unexpected ways on which criteria are chosen and how they are measured.

Within a given state, inferences about the fairness of the maps created through a neutral process might change as the scale of districts varies from massive 700,000-person Congressional districts to, for example, 3,000-person New Hampshire State House districts.  Reformers often point out that U.S. Congressional districts are extremely large relative to districts in other countries that use winner-take-all districts, and a popular reform proposal is to make the U.S. Congress considerably larger by reducing the size of districts.\footnote{See, for instance, ``America Needs a Bigger House,'' New York Times, November 9, 2018.}  Part of the logic of this type of reform is the hope that disproportionalities in the transformation of votes to seats would be reduced if districts were drawn at a smaller scale. In this chapter, however, our finding will be that the Democratic disadvantage in turning votes into seats in Pennsylvania persists at every hypothetical district size, from 4 million people down to just 55,000.

Social scientists and mathematicians are in early stages of understanding the complex interplay of political geography, spatial scale, and statewide partisanship that determine patterns of political representation when districts are drawn without regard for partisanship.  This chapter makes progress by presenting a detailed case study of Pennsylvania.  We choose Pennsylvania in part because in the wake of a recent state court decision, Pennsylvania reformers are in the midst of serious debates about process-oriented versus outcome-oriented reform \cite{nagle2019}.  We make use of modern statistical sampling methods, discussed in more depth in \cite{defordrecombination, book}, to generate large neutral ensembles of possible districting plans in order to study the baseline of representation for each party at a wide range of feasible spatial scales.

Our central conclusion is that given current patterns of political geography in Pennsylvania, purely process-oriented reforms would typically result in the Democratic Party falling significantly short of proportional representation, even when it has a majority of votes, showing that spatial effects overcome the usual ``winner's bonus" \footnote{The ``winner's bonus'' is the established idea in the literature that parties which win the statewide vote should generally have a seat share which exceeds their statewide vote share -- see Bernstein and Walch's chapter in \cite{book} for more.}. We are able to draw inferences about this not only by observing outcomes of very close elections, like the 2016 presidential election, but also by examining statewide elections where Democratic candidates won significant victories, as well as elections in which Republican candidates were victorious.  We are also able to learn subtle lessons about the importance of spatial voting patterns by observing surprisingly different anticipated seat shares associated with elections held on the same day, and with very similar overall partisan vote shares, but with different underlying spatial support patterns.  

Second, we demonstrate that while the scale of districts does affect the baseline for representation, the effect is largely to decrease the variance and not to reduce the gap between expected seat share and the statewide vote share. In other words, the Democrats' geography problem does not simply go away if districts become sufficiently small. In closely contested elections, at no plausible scale of redistricting do our neutral ensembles produce Democratic seat shares that match their vote shares.  

\begin{figure}
\centering
\includegraphics[width=0.7\textwidth]{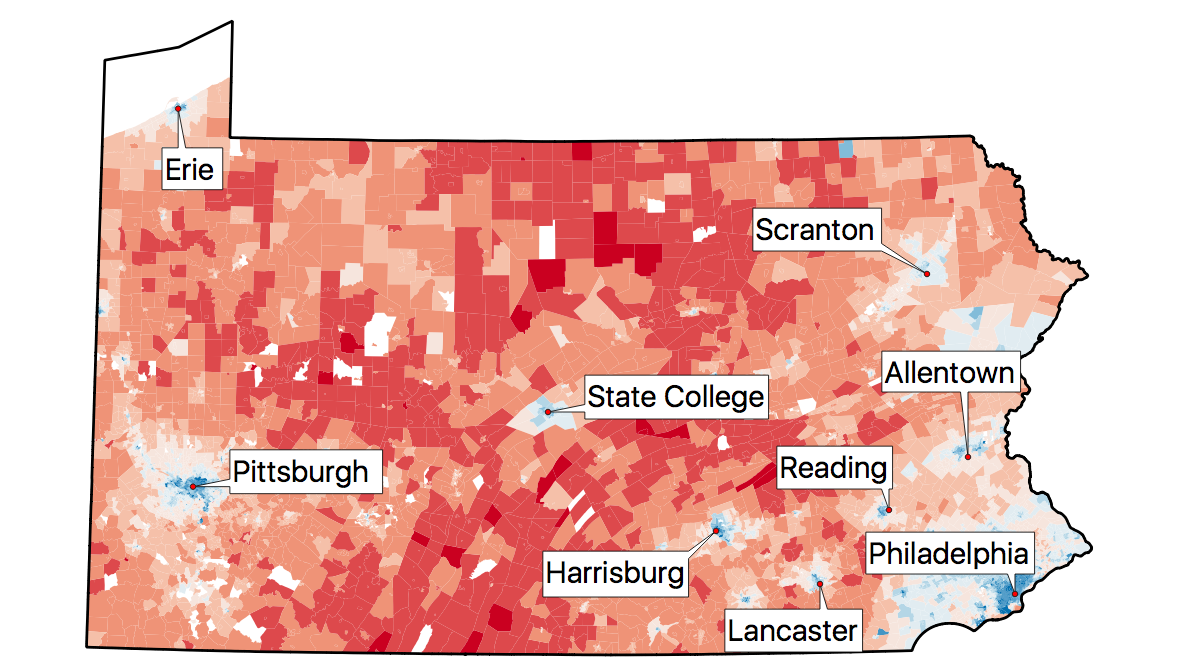}
\caption{Map of Pennsylvania with results from the 2016 Presidential election.}
\label{labelledPA}
\end{figure}

Third, the main reason for choosing Pennsylvania as our case study is that by dividing the state in half and treating Eastern and Western Pennsylvania as two separate states, we are able to gain a better understanding of what exactly lies behind the Republican advantage.  That is, we are able to gain insights by ``modularizing" the problem into two smaller problems.  Pennsylvania gives us the opportunity to examine two very different, and perhaps somewhat representative, patterns of political geography.  Eastern Pennsylvania contains not only a large, extremely Democratic ``primate'' city, but also, due to the geography of coal and the associated 19th-century process of rail-based city formation, a series of smaller Democratic urban agglomerations located in close proximity to one another (see Figure \ref{labelledPA}).  This pattern of smaller, geographically proximate corridors of post-industrial cities that grew up along rail lines in the periphery of larger regional primate cities resembles other early-industrializing states along the Eastern Seaboard.  In Pennsylvania, this pattern is associated with an unambiguous but relatively modest  pattern of Republican advantage in our ensembles of non-partisan redistricting plans.  

Western Pennsylvania, in contrast, contains a single large ``primate'' city that is overwhelmingly Democratic, while smaller Democratic enclaves are few in number and quite isolated from one another.  This pattern of political geography is also found in other states on the Western and Southern fringes of the early 20th century manufacturing core, like Missouri, Tennessee, and Louisiana, which contain relatively large, extremely Democratic cities, but lack a network of smaller, proximate rail-based agglomerations on the order of  Allentown, Easton, and Reading.  

Our analysis demonstrates that relative to the Eastern Pennsylvania pattern of dense industrialized corridors, this Western Pennsylvania structure featuring a single isolated 19th century industrial outpost is associated with a much greater under-representation of 21st century Democrats.  Non-partisan redistricting plans grant Republicans substantially more seats than proportional representation would suggest in Western Pennsylvania, with this phenomenon driving a large part of the overall story of Democratic under-representation in our ensembles of statewide maps.             

We begin with a brief discussion of American political geography and the normative challenge that it creates for a scheme of representation that relies on dividing the states up into winner-take-all districts.  Next, we describe our empirical strategy for generating samples of non-partisan redistricting plans at various spatial scales.  We then explore the key characteristics of our sampled statewide plans, paying special attention to issues of 1) spatial scale and 2) the heterogeneity in statewide partisanship and political geography associated with different statewide Pennsylvania elections. Python code relating to this chapter is available at \url{https://github.com/political-geometry/chapter-3-political-geography}

%%%
%%%
\section{Urban geography and the partisan tilt of neutral redistricting}

Until recently, the debate about redistricting reform in the United States pitted those who believe that redistricting should remain in the hands of legislative majorities against those who believe it should be delegated to either non-partisan or bipartisan commissions.  The prevailing model among the latter group was that of the non-partisan commissions employed in Great Britain, Canada, and Australia, or in the U.S. context, the Iowa process.  All of these prevent those drawing the maps from having access to data on partisanship.  More recently, among those who favor redistricting reform, a new debate has emerged.  Should reformers attempt to stick with some form of party-blind process, or include some measure of anticipated partisan symmetry in the marching orders of commissions?  

This debate has been spurred by a literature that builds on observations of classic British and Australian political geographers \cite{gudgintaylor79, johnston1977, johnston2001}.  Ever since the rise of modern parties of the left in the late 19th and early 20th centuries in the era of labor mobilization in industrialized societies, voters for these parties have been highly concentrated in city centers.  This relative concentration is widely believed to underlie their difficulty in transforming votes into seats.  

In the United States, Democrats today are still quite concentrated in the urban core of cities---large and small---that emerged in the era of rapid industrialization, railroad construction, and labor mobilization.  Much has changed since the Democrats emerged as an urban party in the New Deal era, however, and as new issues have been politicized, from civil rights to abortion to guns and now immigration and globalization, the correlation between population density and Democratic voting has increased substantially, and it has spread from the early industrializing states to the entire country, including the deep South. 

The Democratic Party today often suffers from the same difficulty in transforming votes to seats as that faced by Labor parties in the Commonwealth countries in the early postwar era, and so one may wonder if this is also an effect of urban concentration. The reasons behind the under-performance of the Democrats have been hard to nail down, however, because the era of intense urban-rural polarization coincided with highly visible attempts at partisan gerrymandering.  For good reasons, Americans came to see stark disjunctures between votes and seats as phenomena that could be explained purely by partisan gerrymandering.  However, by drawing a series of alternative neutral maps through a simple automated redistricting algorithm, Chen and Rodden \cite{chenrodden2013} showed that in a number of states, substantial Republican advantage would have emerged even in their samples of non-partisan maps, which the authors attributed to the concentration of Democrats in city centers.  This technique was then used to generate a set of comparison maps that was used in court as part of a lawsuit that led to the invalidation of Florida's Congressional redistricting plan in 2014 \cite{chenrodden2015}. The Republican advantage in neutral redistricting is not universal, we should be careful to note. For some elections in Massachusetts, for example, no map (and hence in particular no process, neutral or otherwise) could have garnered the Republican party a congressional seat despite statewide vote shares of above 30\% \cite{mgggma}. 

Subsequently, a number of scholars have adopted a series of alternative approaches to sampling from the distribution of possible non-partisan plans, often with the goal of contrasting that distribution with the plan drawn by a state legislature in order to challenge it in court \cite{amicus_geographers, amicus_math, cho_talisman, magleby, mattingly, pegden, pegden2017, mgggva, duchinpa}. This technique has now been used to invalidate redistricting plans in state court in Florida, North Carolina, Michigan, and Pennsylvania.  

As the body of research relying on computer-generated redistricting samples evolves and matures, and as the conversation shifts from court challenges to redistricting reform proposals, it is useful to bring these tools back to the original questions about political geography.  The key question remains: what partisan tendencies should we expect in the absence of gerrymandering intent? In other words, what is the impact of the unavoidable consequences of districts on proportionality and other fairness measures, and how does this impact change from election to election and between scales of redistricting?

Although U.S. political geography is always changing, in the current moment, when population density and Democratic voting are correlated at unprecedented levels, the answer to this question appears to lie largely in the size, structure, and arrangement of cities and suburbs relative to their rural surroundings.  A basic problem is that large cities like Philadelphia and Pittsburgh are overwhelmingly Democratic, and neutral redistricting plans will tend to produce overwhelmingly Democratic districts.  On the other hand, ``rural'' districts often encompass not only a large number of Republicans, but also non-trivial clusters of far-flung Democrats in agglomerations like Erie and State College, Pennsylvania that are too small to produce Democratic majorities.  Moreover, districts that are largely exurban will often contain fragments of heavily Democratic parts of inner- and middle-ring suburbs.  As a result, even if their statewide vote shares are similar, Republican candidates tend to win victories by smaller margins than do Democrats, whose votes are inefficiently concentrated in the districts they win.  

Urban concentration's effect on representation is highly dependent on the size, arrangement, and structure of cities as well as the scale at which districts are drawn.  When cities are very large relative to the size of districts, e.g. Philadelphia and Pittsburgh, a non-partisan process would likely create extremely Democratic urban districts.  When cities are too small relative to the size of districts, as with Erie and State College relative to Congressional districts, Democrats are unable to form majorities.  But sometimes the size of a city is better for the representation of Democrats.  For example, some of Pennsylvania's mid-sized cities, like Reading and Bethlehem, are close to the ideal size for producing comfortable but not overwhelming Democratic state Senate Seats.  But at the much smaller scale of State House districts, these cities can produce overwhelming Democratic majorities akin to Philadelphia. What is far from clear, however, is what the effects of changing redistricting scales will be on aggregate: smaller districts may lead to small Democratic towns electing Democratic representatives, but is this effect enough to change the overall seat share in the chamber in question? 

As we demonstrate, it also matters a great deal how these cities are arranged in space.  In addition to Philadelphia, the cities of Eastern Pennsylvania grew up in the late 19th century along a dense web of railroads that were built around the economic geography of coal mining and heavy industry. As a result, Eastern Pennsylvania ended up with a series of small, proximate rail-based industrial towns.  Scranton and Wilkes-Barre blend together along a seam of coal to the North.  Further South, Easton, Bethlehem, and Allentown blend together into a Democratic corridor.  Continuing in a ring around Philadelphia is a series of smaller, extremely Democratic railroad cities including Reading, Lancaster, Harrisburg, and Chester.  At the scale of Congressional districts, these cities are sufficiently proximate to one another, and to some Democratic suburbs of Philadelphia, that they can string together to produce Democratic majorities.  

Another consideration is urban form.  Some 19th-century cities, like Pittsburgh, have a dense and Democratic urban core, and as one moves to the suburbs, the Republican vote share increases rapidly, which generates a highly concentrated Democratic population.  In contrast, the growth of Republican vote shares is slower as one moves from the core to the outer-ring suburbs in Philadelphia, in part because of the locations of high-technology employers, colleges, and universities, whose employees have become important parts of the Democratic coalition.  

\begin{figure}
    \centering
    \includegraphics[width=\textwidth]{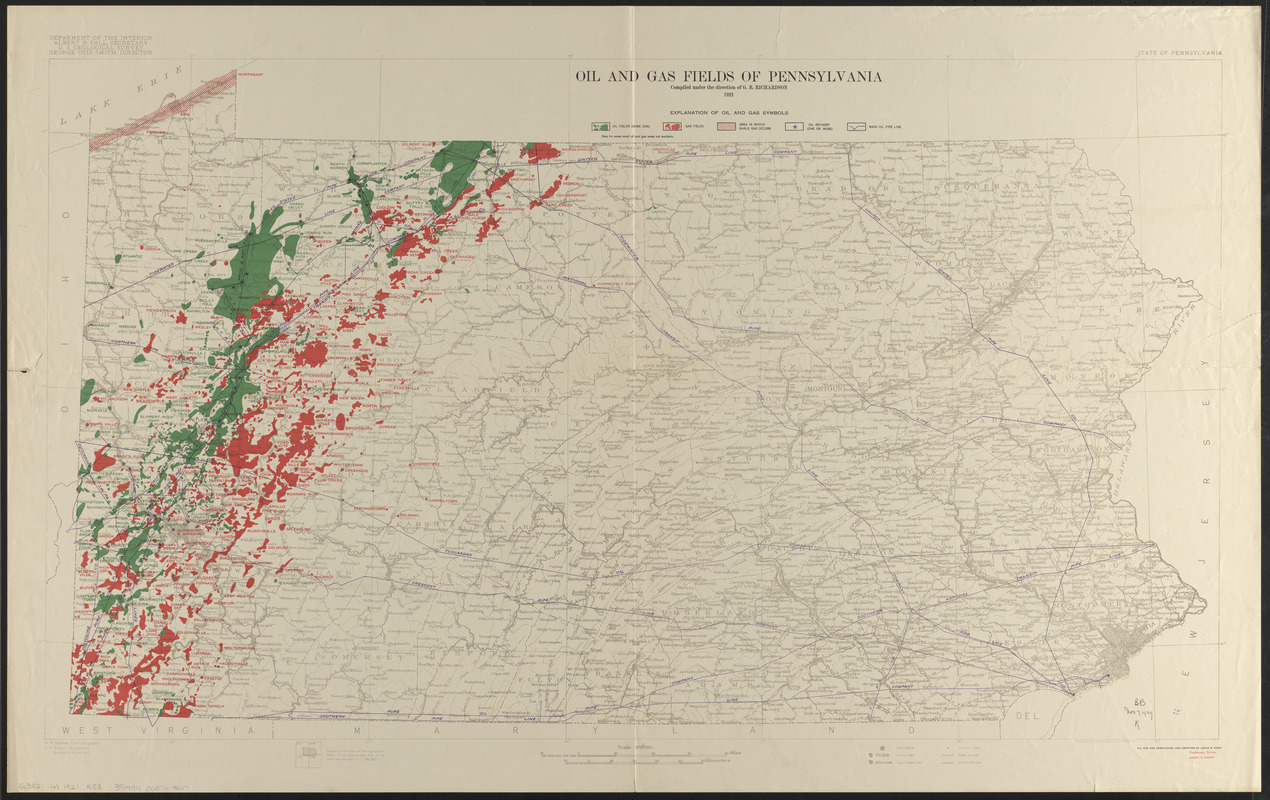}
    \caption{Oil and gas fields of Pennsylvania, from 1921. Taken from the Norman B. Leventhal Map Center Collection.}
    \label{fig:oil_gas_pa}
\end{figure}

\begin{figure}
    \centering
    \includegraphics[width=\textwidth]{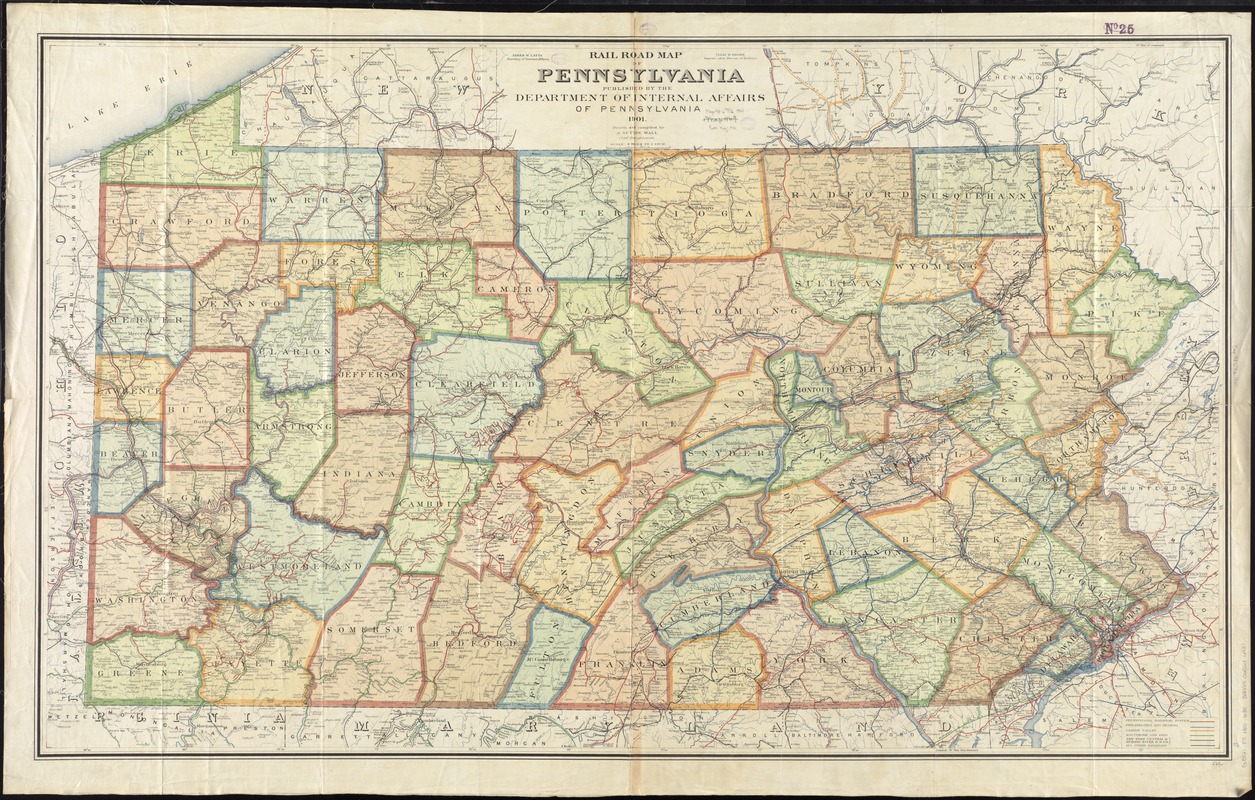}
    \caption{Railroad map of Pennsylvania from 1901. Taken from the Norman B. Leventhal Map Center Collection.}
    \label{fig:pa_1901_railroads}
\end{figure}

Even more distinct are cities like Orlando or Phoenix, where there is no 19th century core, and Democrats and Republicans are interspersed in a sprawling poly-centric metropolis.  Moreover, there are parts of the United States with important pockets of rural support for Democrats, including African-American communities in the South, tribal lands, and communities with a history of mining.      

With the continued focus on the possible detrimental effects of urban concentration on representation, we should be careful not to exclude the possibility that in some cases this concentration may \emph{help} a party in the transformation of votes to seats, particularly in states where that party typically expects to lose the overall popular vote.  Indeed, the worst case scenario for a losing party is to have its votes perfectly evenly distributed in space, a scenario that would cause it to lose every single district.  This sounds far-fetched, but the situation is Massachusetts is not so far off \cite{mgggma}: the Republicans are unable to gain even one seat in some cases precisely because they are too diluted, not because they are too concentrated.

In short, the location of Democrats in cities is not a sufficient condition to produce a Republican advantage in neutral redistricting plans, and the extent of that advantage, when it exists, is potentially a function of political geography, the scale at which districts are drawn, and the overall level of support for the party.  Our goal in this chapter is to take a close look at Pennsylvania, varying the spatial scale of which districts are drawn, exploring variation in overall vote shares by drawing on a diverse set of recent statewide elections, and exploring the role of heterogeneous urban structure by contrasting the Eastern and Western parts of the state.

\section{Sampling Pennsylvania redistricting plans at different spatial scales}

\subsection*{Method}

We seek to understand the general properties of the universe of all redistricting plans for Pennsylvania. No computer could ever enumerate every possible redistricting plan, but we will use sampling methods that allow us to effectively sample from this vast space in a mathematically rigorous way. The algorithm we will use is the so-called recombination (or {\sf ReCom}) Markov chain algorithm which is implemented in the freely available GerryChain software developed at MGGG and which is discussed  in \cite{defordrecombination, book}. Beginning with a randomly generated ``seed'' plan, the algorithm merges and redivides two adjacent districts at every step, resulting in a large ensemble of randomly generated plans, all drawn without the use of any partisan data. In each case, districts are built out of fixed geographic units: precincts for low numbers of districts and census blocks for higher numbers (in order to make population-balanced plans feasible).\footnote{Some intermediate scales of redistricting were studied with both precincts and blocks, with the two methods producing almost identical results.}

Typically, this kind of algorithm is used to generate a large ensemble of districting plans with a fixed number of districts. This is because one is usually interested in studying the districts for a particular level of government (for example, U.S. Congressional districts, of which there are eighteen in Pennsylvania). For our purposes, we want to study redistricting plans with a variety of different numbers of districts. We thus run the algorithm multiple times (once for each number of districts) to generate many ensembles of 50,000 plans each. For a plan in any one of these ensembles, we can choose a past election and use the vote data to determine how many districts the Democrats would have won using that election and that redistricting plan;\footnote{Since election data is only reported at the precinct level we prorate it down to blocks based on voting age population when needed.} dividing this number by the total number of districts in the plan gives us the \emph{Democratic seat share}. We chose nine elections to base our analysis on: the presidential elections from 2012 and 2016 (PRES12 and PRES16), the U.S. Senate elections from 2010, 2012 and 2016 (SEN12 and SEN16), the Attorney General elections from 2012 and 2016 (ATG12 and ATG16) and the Gubernatorial elections from 2010 and 2014 (GOV10 and GOV14). In each case we just treated the election as a head-to-head election between Democrats and Republicans, ignoring any third party candidates.

\subsection*{Results}

\begin{figure}
\centering
\includegraphics[width=\textwidth]{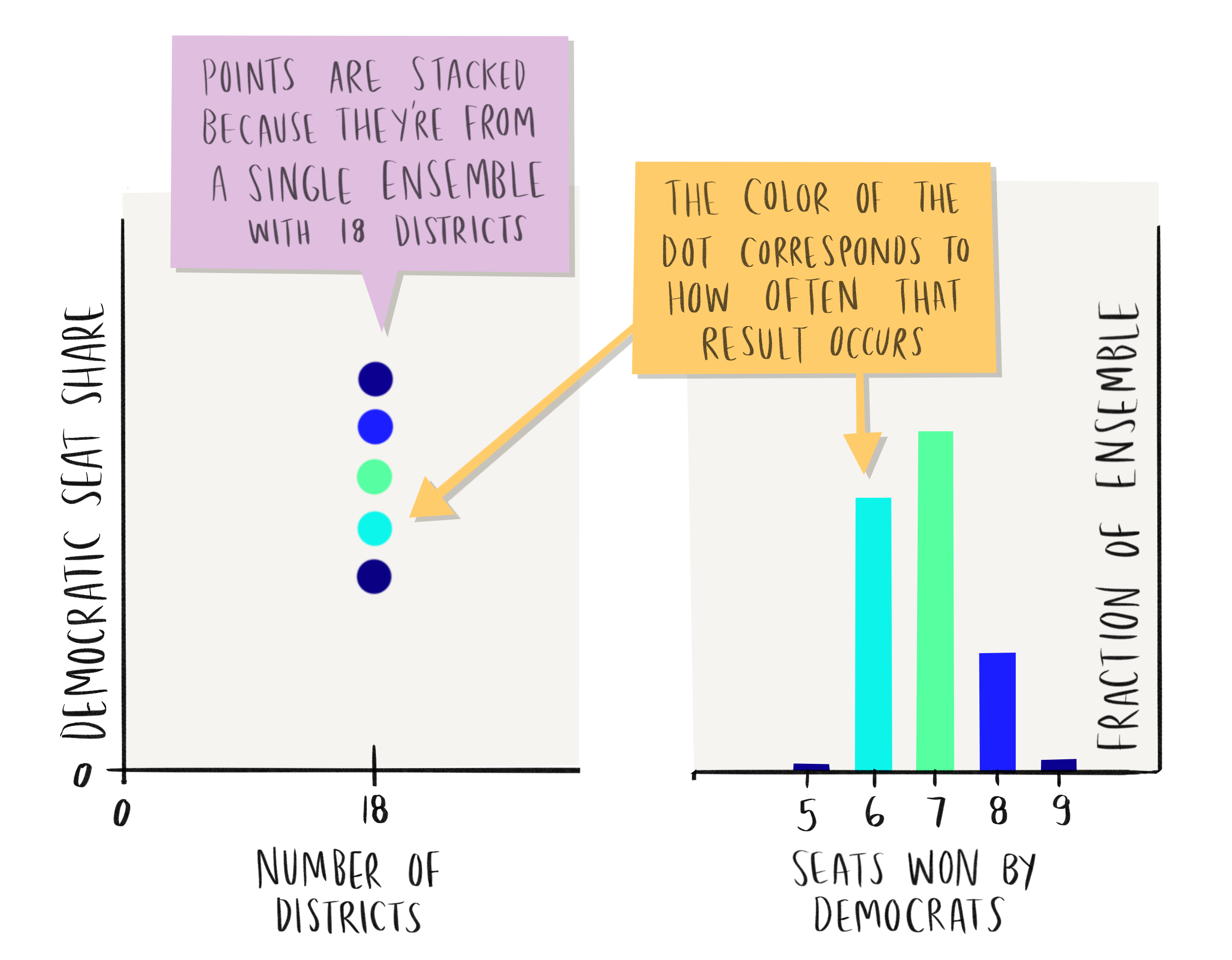}
\caption{An illustration of how the plots in this section are constructed. On the right is a histogram of seats outcomes for an ensemble of 18-district plans, using PRES16 vote data. Each bar of this histogram is converted to a dot in the left hand plot, with the brightness of the color indicating the bar height, the $x$-value denoting the number of districts (in this case, 18) and the $y$ axis indicating the fraction of seats won.}
\label{toyexample}

\end{figure}

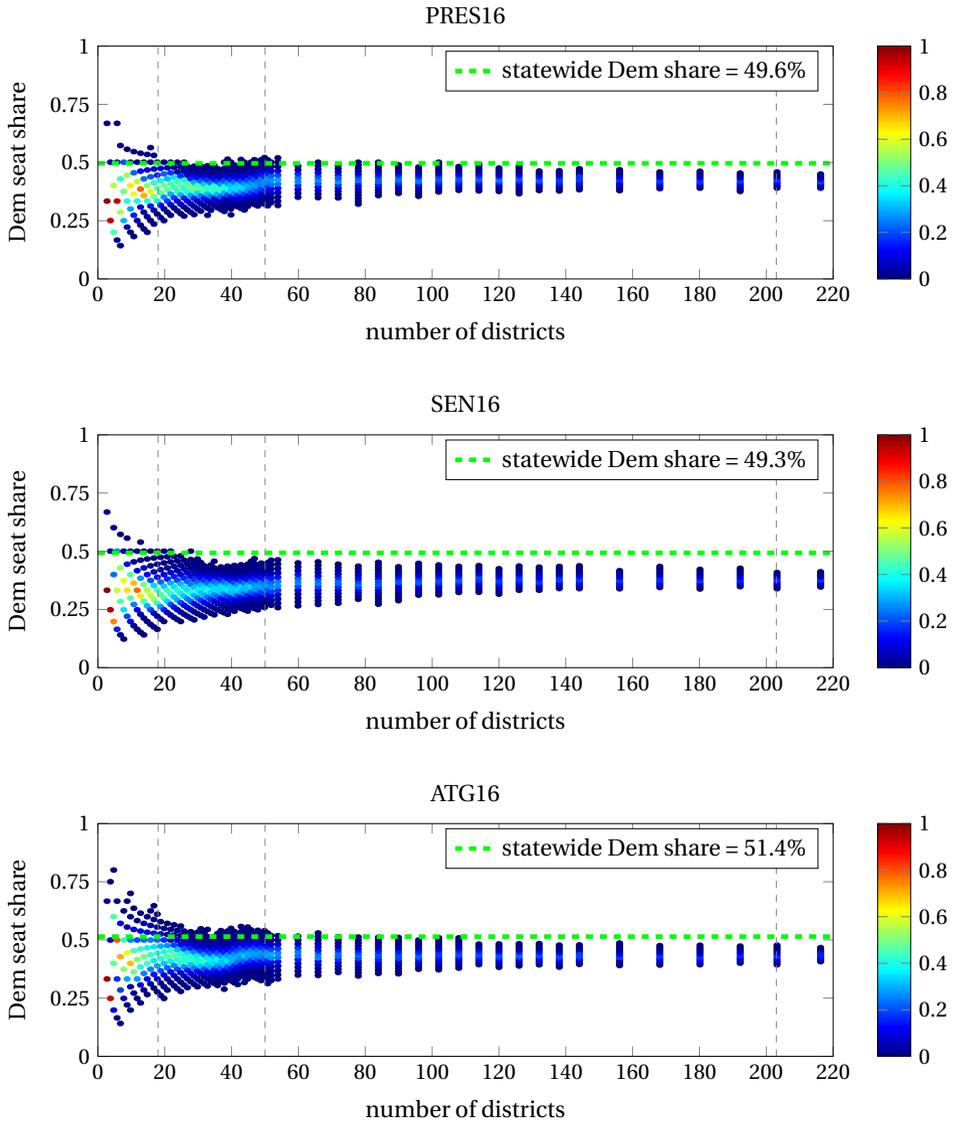
\begin{figure}
\begin{subfigure}{\textwidth}
\centering
\resizebox{\columnwidth}{!}{%
\begin{tikzpicture}
\begin{axis}[
title=PRES16,
xmin=0,xmax=220,
ymin=0, ymax=1,
axis on top,
width=\textwidth, height=0.4\textwidth,
ytick={0,0.25,0.5,0.75,1},
colormap/jet, mark size=1,
ylabel = Dem seat share,
xlabel = number of districts,
point meta min = 0,
point meta min = 1,
colorbar
]
\addplot[forget plot] graphics[xmin=0,xmax=220, ymin=0, ymax=1] {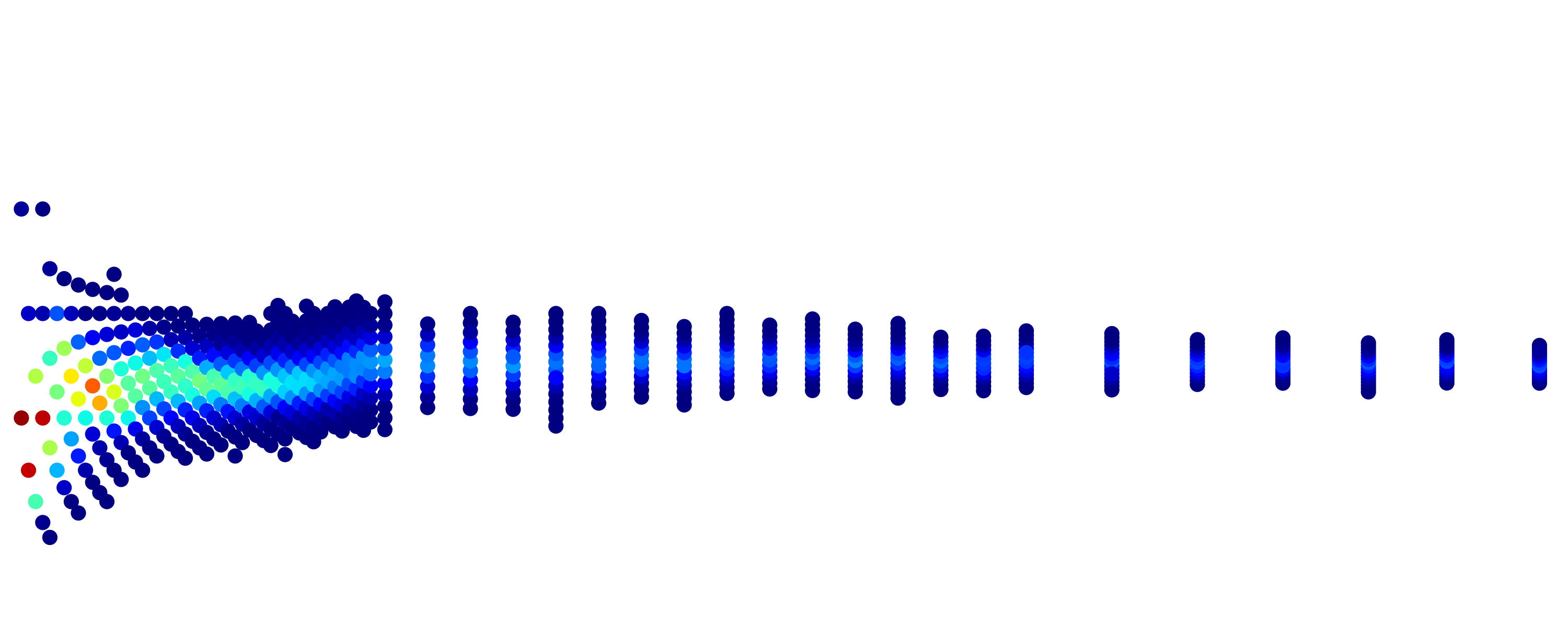};
\addplot[green, dashed, line legend, line width=2pt] coordinates{(0,0.49646) (220,0.49646)};

\addplot[gray, forget plot, dashed] coordinates{(18,0) (18,1)};
\addplot[gray, forget plot, dashed] coordinates{(50,0) (50,1)};
\addplot[gray, forget plot, dashed] coordinates{(203,0) (203,1)};

\legend{statewide Dem share = 49.6\%}
\end{axis}
\end{tikzpicture}
}
\caption*{}
\end{subfigure}

\begin{subfigure}{\textwidth}
\centering
\resizebox{\columnwidth}{!}{%
\begin{tikzpicture}
\begin{axis}[
title=SEN16,
xmin=0,xmax=220,
ymin=0, ymax=1,
axis on top,
width=\textwidth, height=0.4\textwidth,
ytick={0,0.25,0.5,0.75,1},
colormap/jet, mark size=1,
ylabel = Dem seat share,
xlabel = number of districts,
point meta min = 0,
point meta min = 1,
colorbar
]
\addplot[forget plot] graphics[xmin=0,xmax=220, ymin=0, ymax=1] {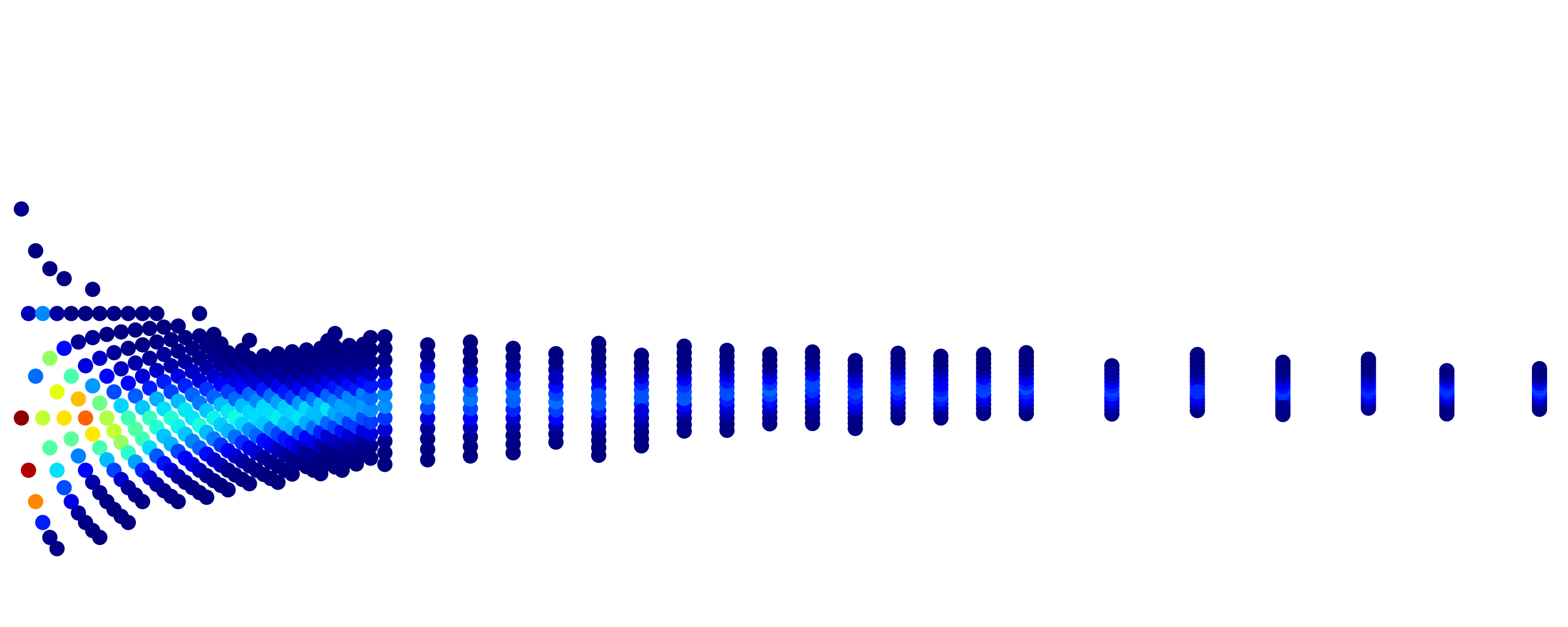};
\addplot[green, dashed, line legend, line width=2pt] coordinates{(0,0.492846) (220,0.492846)};

\addplot[gray, forget plot, dashed] coordinates{(18,0) (18,1)};
\addplot[gray, forget plot, dashed] coordinates{(50,0) (50,1)};
\addplot[gray, forget plot, dashed] coordinates{(203,0) (203,1)};

\legend{statewide Dem share = 49.3\%}
\end{axis}
\end{tikzpicture}
}
\caption*{}
\end{subfigure}

\begin{subfigure}{\textwidth}
\centering
\resizebox{\columnwidth}{!}{%
\begin{tikzpicture}
\begin{axis}[
title=ATG16,
xmin=0,xmax=220,
ymin=0, ymax=1,
axis on top,
width=\textwidth, height=0.4\textwidth,
ytick={0,0.25,0.5,0.75,1},
colormap/jet, mark size=1,
ylabel = Dem seat share,
xlabel = number of districts,
point meta min = 0,
point meta min = 1,
colorbar
]
\addplot[forget plot] graphics[xmin=0,xmax=220, ymin=0, ymax=1] {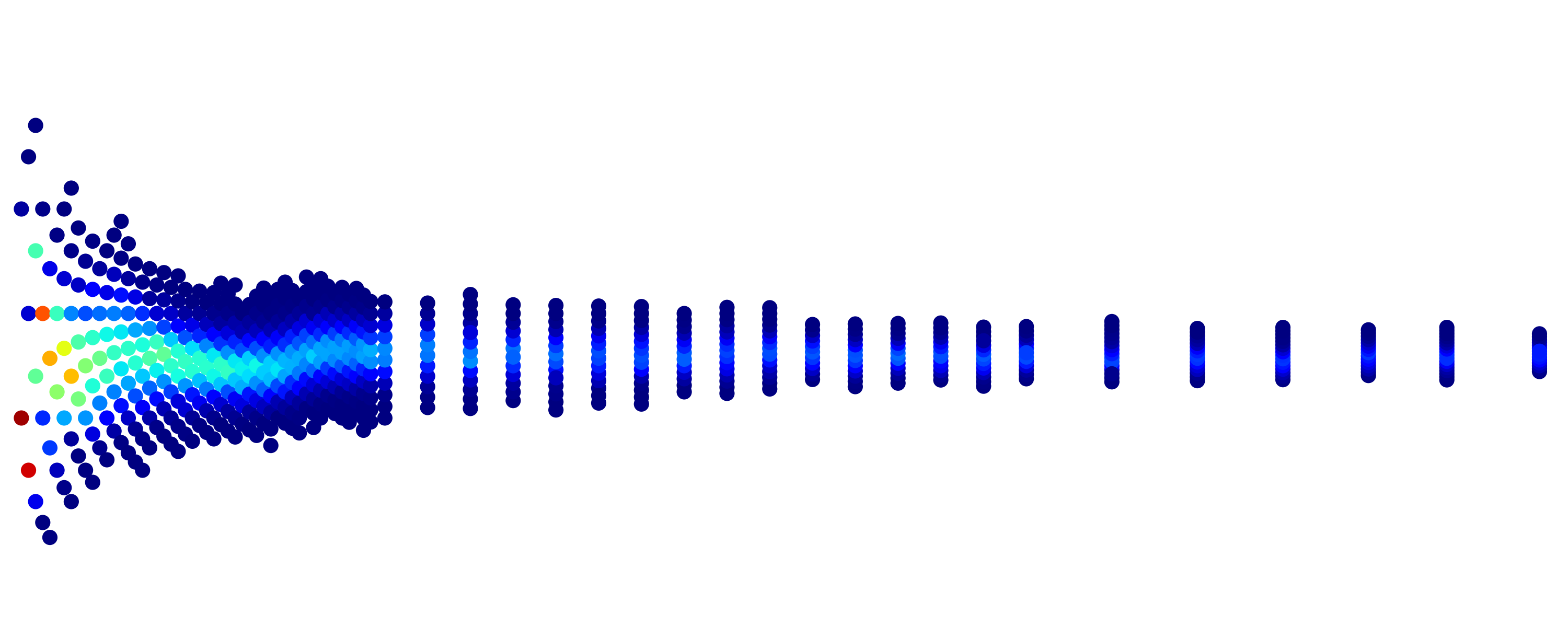};
\addplot[green, dashed, line legend, line width=2pt] coordinates{(0,0.5143) (220,0.5143)};

\addplot[gray, forget plot, dashed] coordinates{(18,0) (18,1)};
\addplot[gray, forget plot, dashed] coordinates{(50,0) (50,1)};
\addplot[gray, forget plot, dashed] coordinates{(203,0) (203,1)};

\legend{statewide Dem share = 51.4\%}
\end{axis}
\end{tikzpicture}
}
\caption*{}
\end{subfigure}
\caption{Democratic seat shares in neutral ensembles at various redistricting scales, 2016 elections}
\label{scale16}
\end{figure}

\begin{figure}
\begin{subfigure}{\textwidth}
\centering
\resizebox{\columnwidth}{!}{%
\begin{tikzpicture}
\begin{axis}[
title=PRES12,
xmin=0,xmax=220,
ymin=0, ymax=1,
axis on top,
width=\textwidth, height=0.4\textwidth,
ytick={0,0.25,0.5,0.75,1},
colormap/jet, mark size=1,
ylabel = Dem seat share,
xlabel = number of districts,
point meta min = 0,
point meta min = 1,
colorbar
]
\addplot[forget plot] graphics[xmin=0,xmax=220, ymin=0, ymax=1] {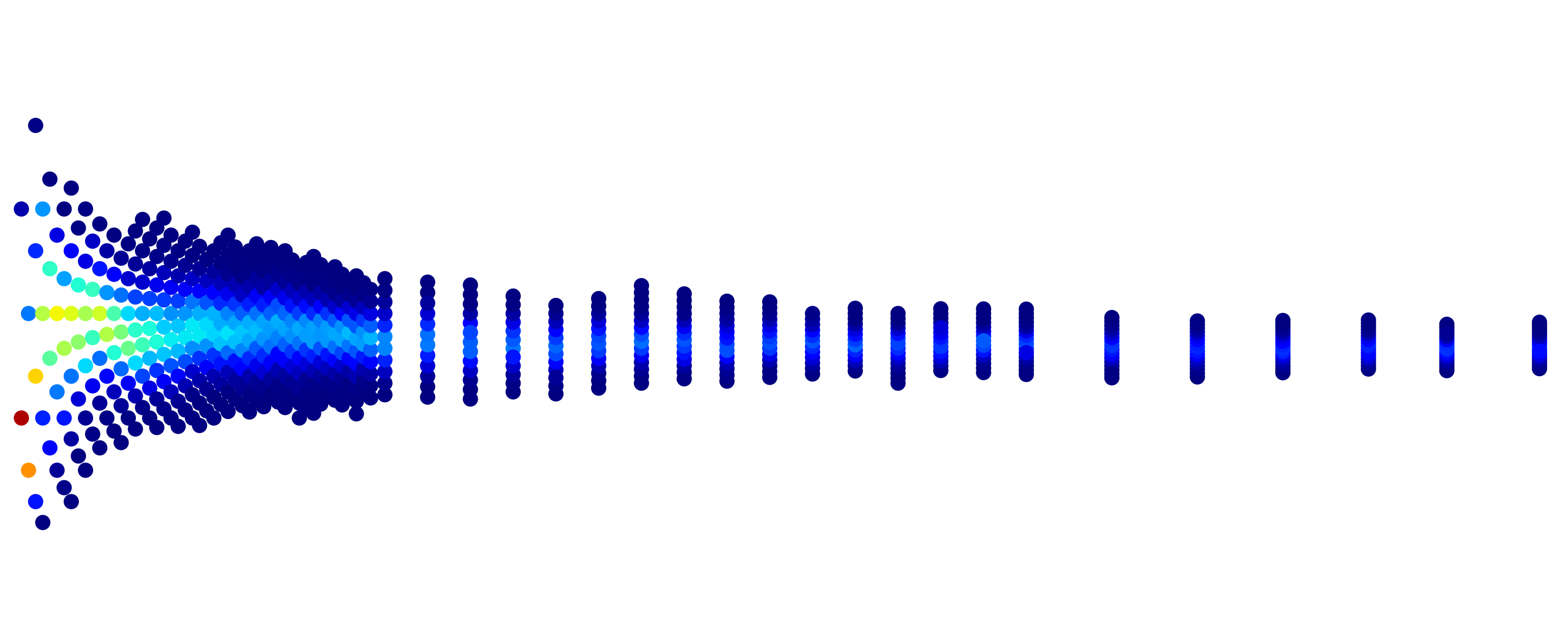};
\addplot[green, dashed, line legend, line width=2pt] coordinates{(0,0.5271) (220,0.5271)};

\addplot[gray, forget plot, dashed] coordinates{(18,0) (18,1)};
\addplot[gray, forget plot, dashed] coordinates{(50,0) (50,1)};
\addplot[gray, forget plot, dashed] coordinates{(203,0) (203,1)};

\legend{statewide Dem share = 52.7\%}
\end{axis}
\end{tikzpicture}
}
\caption*{}
\end{subfigure}

\begin{subfigure}{\textwidth}
\centering
\resizebox{\columnwidth}{!}{%
\begin{tikzpicture}
\begin{axis}[
title=SEN12,
xmin=0,xmax=220,
ymin=0, ymax=1,
axis on top,
width=\textwidth, height=0.4\textwidth,
ytick={0,0.25,0.5,0.75,1},
colormap/jet, mark size=1,
ylabel = Dem seat share,
xlabel = number of districts,
point meta min = 0,
point meta min = 1,
colorbar
]
\addplot[forget plot] graphics[xmin=0,xmax=220, ymin=0, ymax=1] {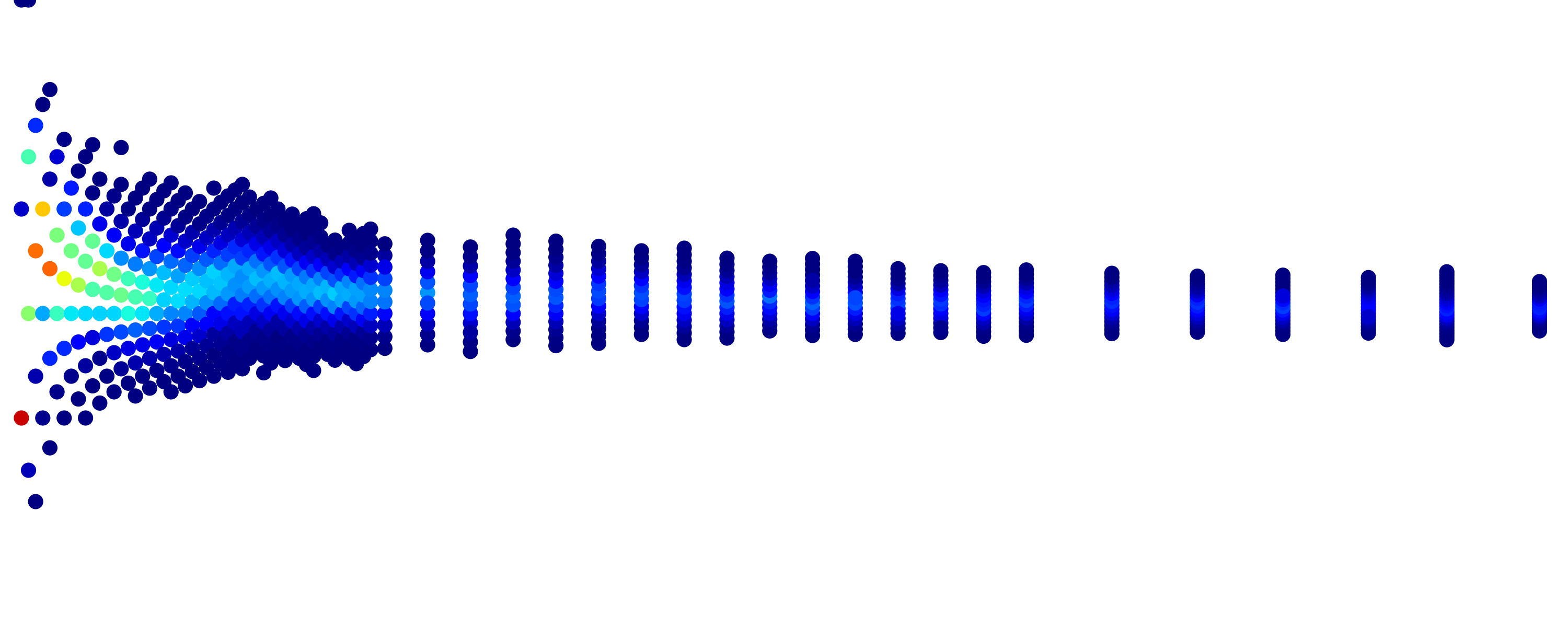};
\addplot[green, dashed, line legend, line width=2pt] coordinates{(0,0.5456) (220,0.5456)};

\addplot[gray, forget plot, dashed] coordinates{(18,0) (18,1)};
\addplot[gray, forget plot, dashed] coordinates{(50,0) (50,1)};
\addplot[gray, forget plot, dashed] coordinates{(203,0) (203,1)};

\legend{statewide Dem share = 54.6\%}
\end{axis}
\end{tikzpicture}
}
\caption*{}
\end{subfigure}

\begin{subfigure}{\textwidth}
\centering
\resizebox{\columnwidth}{!}{%
\begin{tikzpicture}
\begin{axis}[
title=ATG12,
xmin=0,xmax=220,
ymin=0, ymax=1,
axis on top,
width=\textwidth, height=0.4\textwidth,
ytick={0,0.25,0.5,0.75,1},
colormap/jet, mark size=1,
ylabel = Dem seat share,
xlabel = number of districts,
point meta min = 0,
point meta min = 1,
colorbar
]
\addplot[forget plot] graphics[xmin=0,xmax=220, ymin=0, ymax=1] {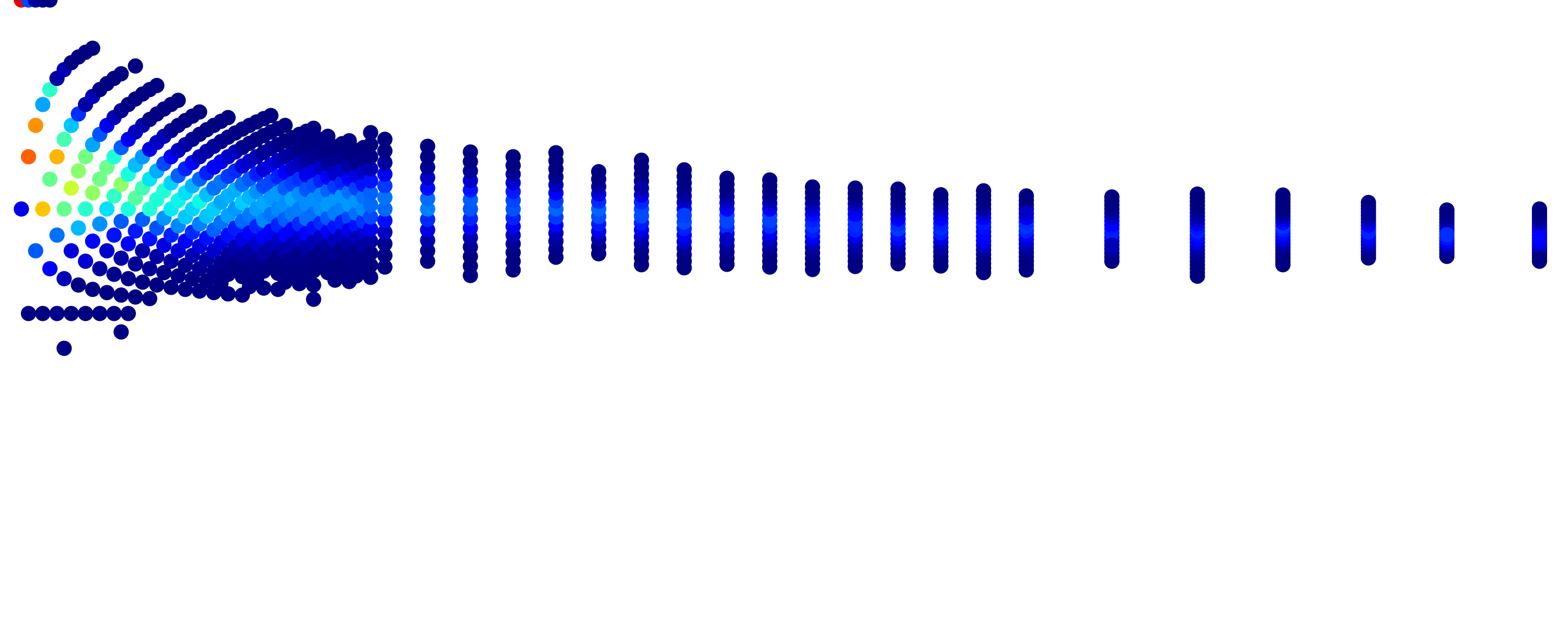};
\addplot[green, dashed, line legend, line width=2pt] coordinates{(0,0.5742) (220,0.5742)};

\addplot[gray, forget plot, dashed] coordinates{(18,0) (18,1)};
\addplot[gray, forget plot, dashed] coordinates{(50,0) (50,1)};
\addplot[gray, forget plot, dashed] coordinates{(203,0) (203,1)};

\legend{statewide Dem share = 57.4\%}
\end{axis}
\end{tikzpicture}
}
\caption*{}
\end{subfigure}
\caption{Democratic seat shares in neutral ensembles at various redistricting scales, 2012 elections.}
\label{scale12}
\end{figure}

\begin{figure}
\begin{subfigure}{\textwidth}
\centering
\resizebox{\columnwidth}{!}{%
\begin{tikzpicture}
\begin{axis}[
title=GOV14,
xmin=0,xmax=220,
ymin=0, ymax=1,
axis on top,
width=\textwidth, height=0.4\textwidth,
ytick={0,0.25,0.5,0.75,1},
colormap/jet, mark size=1,
ylabel = Dem seat share,
xlabel = number of districts,
point meta min = 0,
point meta min = 1,
colorbar
]
\addplot[forget plot] graphics[xmin=0,xmax=220, ymin=0, ymax=1] {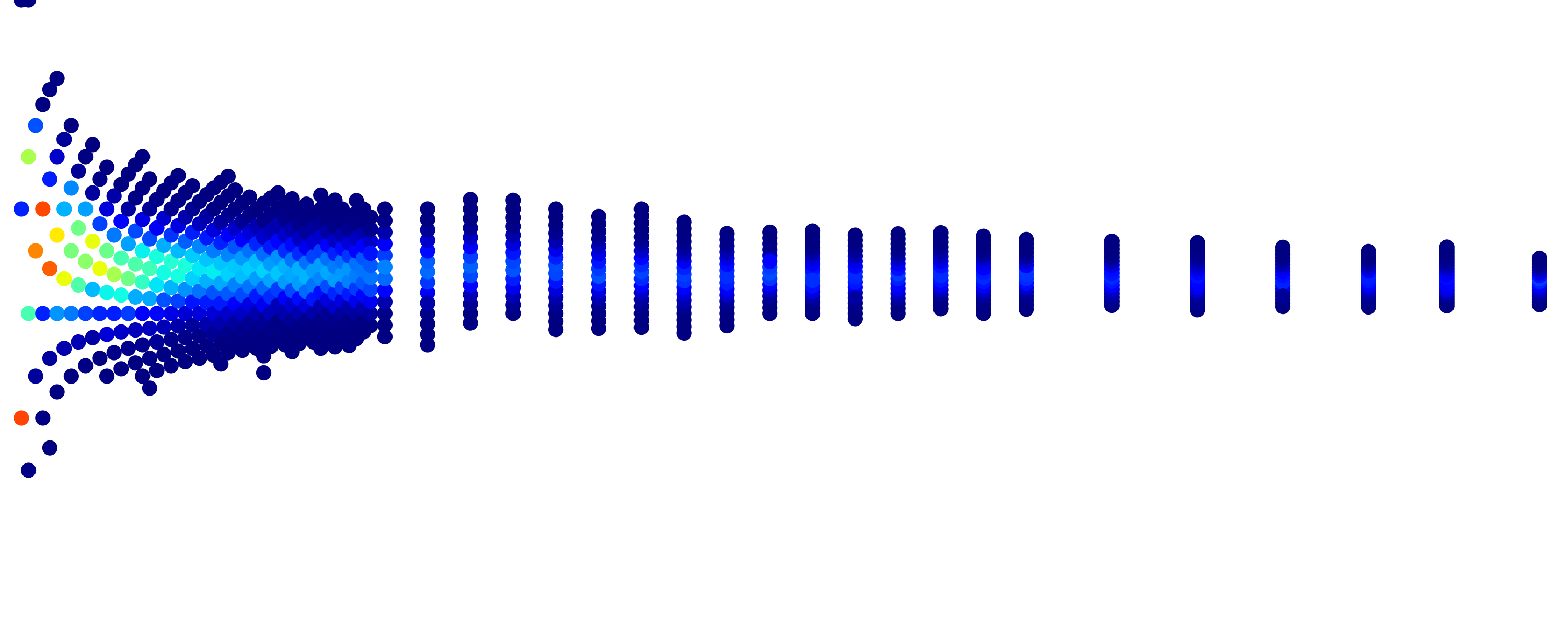};
\addplot[green, dashed, line legend, line width=2pt] coordinates{(0,0.5478) (220,0.5478)};

\addplot[gray, forget plot, dashed] coordinates{(18,0) (18,1)};
\addplot[gray, forget plot, dashed] coordinates{(50,0) (50,1)};
\addplot[gray, forget plot, dashed] coordinates{(203,0) (203,1)};

\legend{statewide Dem share = 54.8\%}
\end{axis}
\end{tikzpicture}
}
\caption*{}
\end{subfigure}

\begin{subfigure}{\textwidth}
\centering
\resizebox{\columnwidth}{!}{%
\begin{tikzpicture}
\begin{axis}[
title=GOV10,
xmin=0,xmax=220,
ymin=0, ymax=1,
axis on top,
width=\textwidth, height=0.4\textwidth,
ytick={0,0.25,0.5,0.75,1},
colormap/jet, mark size=1,
ylabel = Dem seat share,
xlabel = number of districts,
point meta min = 0,
point meta min = 1,
colorbar
]
\addplot[forget plot] graphics[xmin=0,xmax=220, ymin=0, ymax=1] {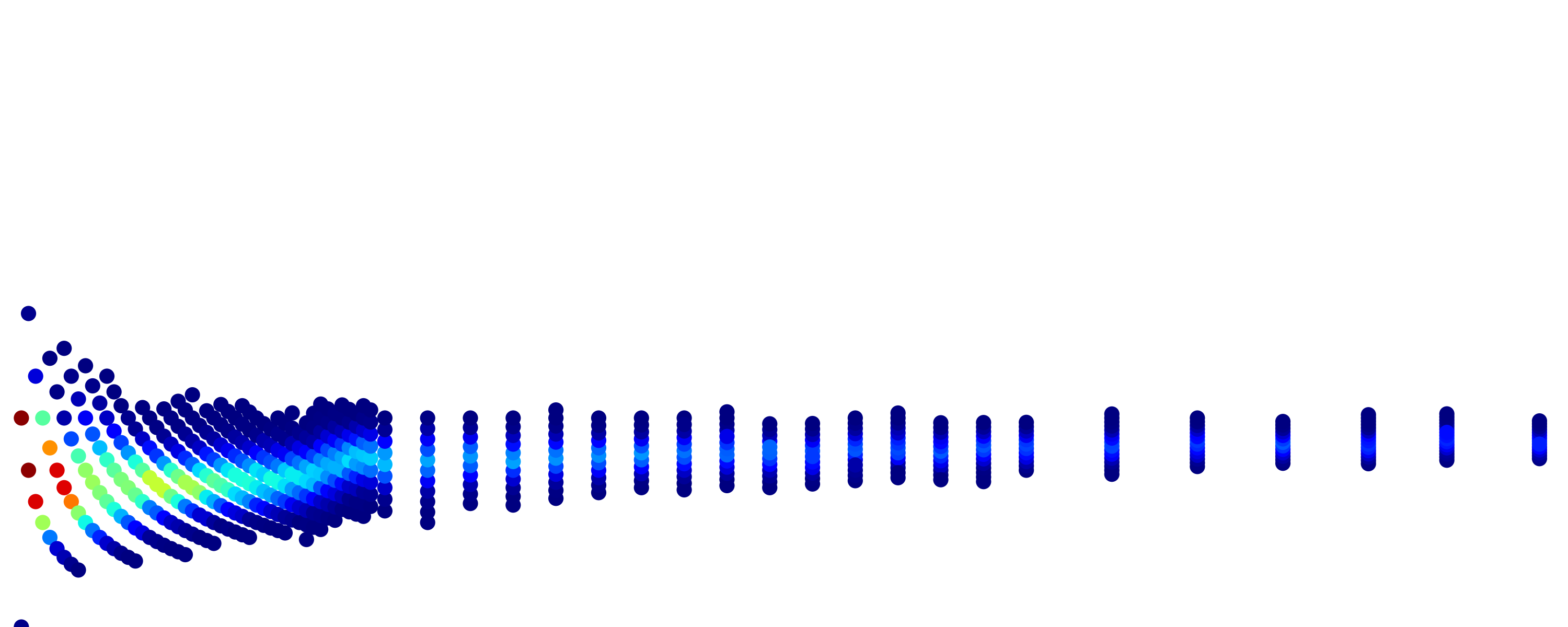};
\addplot[green, dashed, line legend, line width=2pt] coordinates{(0,0.4548) (220,0.4548)};

\addplot[gray, forget plot, dashed] coordinates{(18,0) (18,1)};
\addplot[gray, forget plot, dashed] coordinates{(50,0) (50,1)};
\addplot[gray, forget plot, dashed] coordinates{(203,0) (203,1)};

\legend{statewide Dem share = 45.5\%}
\end{axis}
\end{tikzpicture}
}
\caption*{}
\end{subfigure}

\begin{subfigure}{\textwidth}
\centering
\resizebox{\columnwidth}{!}{%
\begin{tikzpicture}
\begin{axis}[
title=SEN10,
xmin=0,xmax=220,
ymin=0, ymax=1,
axis on top,
width=\textwidth, height=0.4\textwidth,
ytick={0,0.25,0.5,0.75,1},
colormap/jet, mark size=1,
ylabel = Dem seat share,
xlabel = number of districts,
point meta min = 0,
point meta min = 1,
colorbar
]
\addplot[forget plot] graphics[xmin=0,xmax=220, ymin=0, ymax=1] {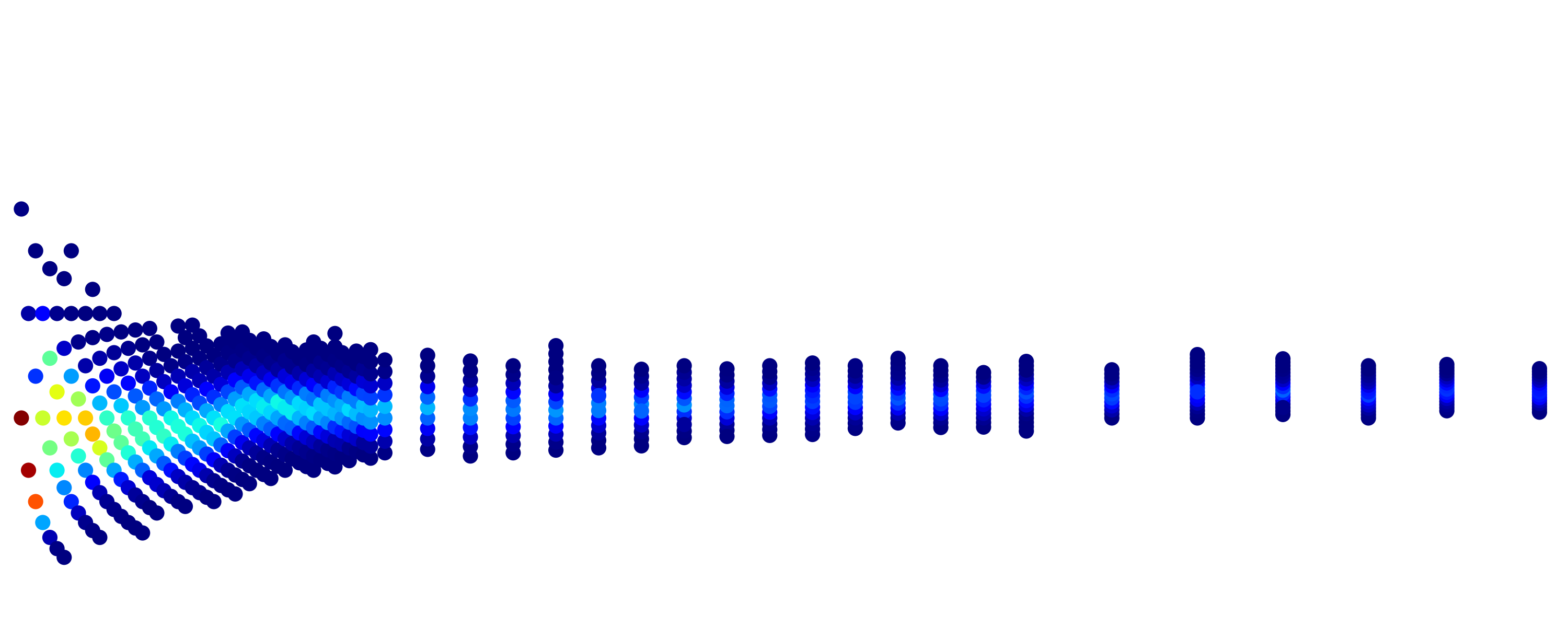};
\addplot[green, dashed, line legend, line width=2pt] coordinates{(0,0.4895) (220,0.4895)};

\addplot[gray, forget plot, dashed] coordinates{(18,0) (18,1)};
\addplot[gray, forget plot, dashed] coordinates{(50,0) (50,1)};
\addplot[gray, forget plot, dashed] coordinates{(203,0) (203,1)};

\legend{statewide Dem share = 49.0\%}
\end{axis}
\end{tikzpicture}
}
\caption*{}
\end{subfigure}
\caption{Democratic seat shares in neutral ensembles at various redistricting scales, other elections.}
\label{scaleother}
\end{figure}
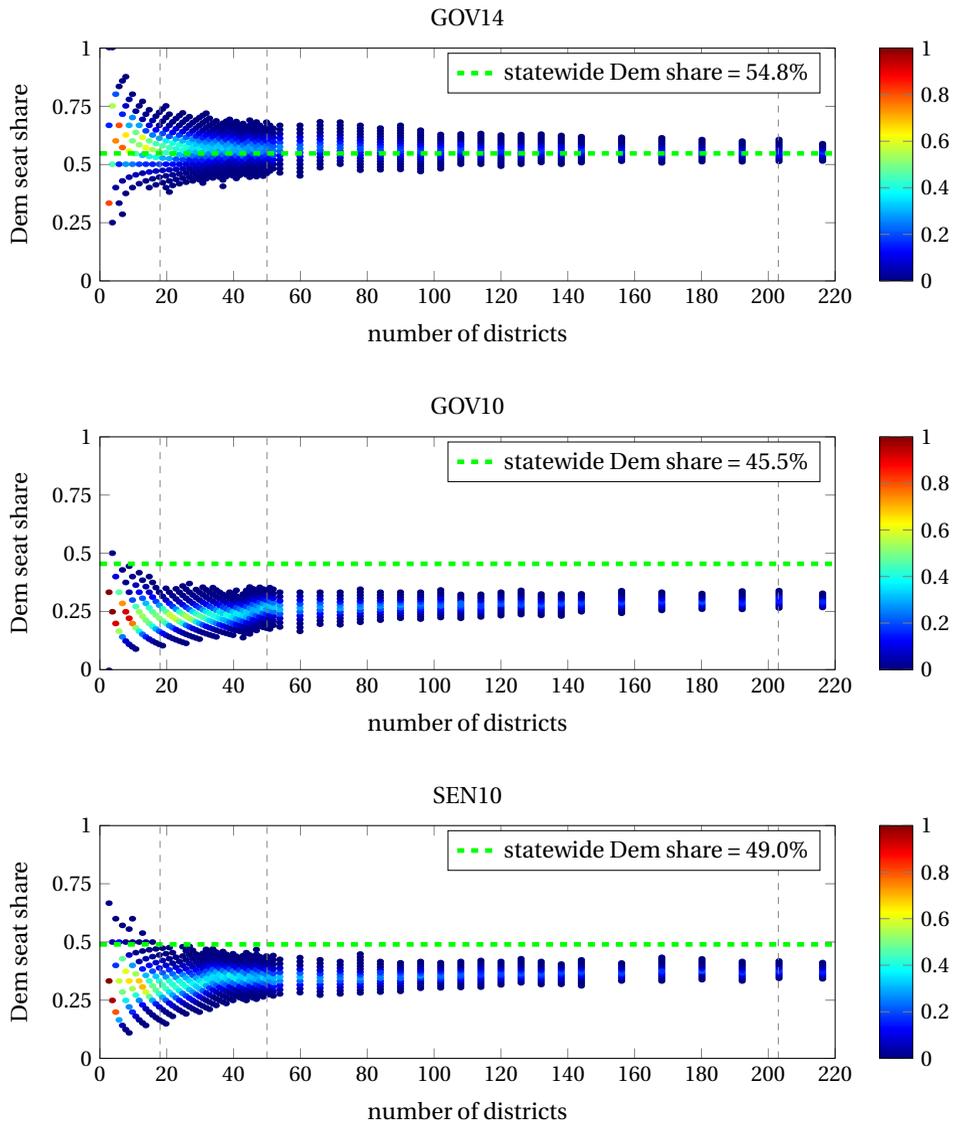

Figures \ref{scale16}, \ref{scale12} and \ref{scaleother} summarize the results of this multi-ensemble analysis, with each plot indicating a different choice of election. Our goal with these plots is to indicate the frequency of different seats outcomes at every scale. In order to make the mode of representation clear, first consider the plots in Figure \ref{toyexample}, where we have isolated just one scale, 18 districts, and where we use PRES16 vote data. On the right is a histogram indicating the seats outcomes for Democrats in the ensemble. This histogram is represented vertically by the colored dots on the left hand plot: dark blue represents a low fraction, with the fraction increasing through light blue (and later to yellow and to red). Notice that we have also switched from number of seats won to fraction of seats won so that we can later compare multiple scales.

% \begin{figure}
% \centering
% \begin{subfigure}{0.4\textwidth}
% \centering
% \resizebox{\columnwidth}{!}{%
% \begin{tikzpicture}
% \begin{axis}[
% width=\textwidth, height=\textwidth,
% ylabel = Democratic seat share,
% xlabel = number of districts,
% ]
% \addplot graphics [xmin=0,xmax=40,ymin=0,ymax=1] {RoddenWeighill/XY_only18.png};
% \end{axis}
% \end{tikzpicture}
% }
% \end{subfigure}
% \begin{subfigure}{0.4\textwidth}
% \centering
% \resizebox{\columnwidth}{!}{%
% \begin{tikzpicture}
% \begin{axis}[
% width=\textwidth, height=\textwidth,
% xtick={5,6,7,8,9},
% ylabel = fraction of ensemble,
% xlabel = Democratic seats won,
% yticklabel pos=right, ylabel near ticks
% ]
% \addplot graphics [xmin=4,xmax=10,ymin=0,ymax=0.5] {RoddenWeighill/hist_18.png};
% \end{axis}
% \end{tikzpicture}
% }
% \end{subfigure}
% \caption{An illustration of how the plots in this section are constructed. On the right is a histogram of seats outcomes for an ensemble of 18-district plans, using PRES16 vote data. Each bar of this histogram is converted to a dot in the left hand plot, with the brightness of the color indicating the bar height, the $x$-value denoting the number of districts (in this case, 18) and the $y$ axis indicating the fraction of seats won.}
% \label{toyexample}
% \end{figure}

In Figures \ref{scale16}, \ref{scale12} and \ref{scaleother}, we employ this scheme to show the seats outcome for all scales at once. Each dot represents a choice of number of districts and a Democratic seat share value. The dot is colored based on what fraction of the plans with that many districts had that Democratic seat share, just as in Figure \ref{toyexample}. Lighter colors therefore represent more common Democratic seat share outcomes for that particular number of districts. The green line represents the statewide vote share achieved by the Democrats in the specified election. Finally, even though we are including hypothetical plans with a large range of numbers of districts, we also indicate three real-world districting scales by dotted lines: U.S. Congress (18 districts), PA state Senate (50 districts) and PA state House (203 districts). 

%%%%%%%%%%%%%%%%%%%%%%%%2016%%%%%%%%%%%%%
%%%%%%%%%%%%

%%%%%%%%%%%%

\subsection*{Discussion}
The most striking conclusion to be drawn from the plots in this section is that the Democrats underperform proportional representation in seven of the nine elections considered, with the only exceptions being the ATG12 and GOV14 elections.  In those two exceptional elections, Democratic candidates performed unusually well, with 57\% and 55\% of the statewide vote respectively.  For elections with relatively even statewide splits between the two parties, like those in 2016, the neutral ensembles showed substantial tendencies to award more seats to the Republican Party. 

Moreover, the Democratic under-performance was more or less unaltered by changing district scales.  If anything, when the Democratic statewide vote share was relatively low, as in GOV10 and GOV14, the Democratic seat share increased very slightly as the scale of districts became smaller.  But when Democrats performed well, as in SEN12 and ATG12, their seat share declined as the scale of districts became smaller. This suggests that a general mismatch between smaller Democratic urban centers and particular districts sizes (for example, Congressional districts) cannot be the only reason for the Democrats' disadvantage, if it plays a role at all. Indeed, these experiments suggest the absence of any significant scale effects.  

The only clear pattern related to the scale of districts in these graphs is the much wider range of seat shares produced by the neutral ensembles when districts are larger (on the left-hand side of the graphs).  The range of outcomes produced in the neutral ensembles narrows considerably as the scale of districts becomes increasingly fine-grained.  Let us focus on the very hotly contested 2016 races, all of which were very close to an even split between the two parties, and where one might expect that a ``fair'' districting plan would produce a roughly similar number of Democratic and Republican seats.  Imagine that a redistricting commission or special master was tasked with the job of randomly selecting a plan from the ensembles.  This would likely lead to a rather large Republican advantage of roughly similar size, whether the plan was for Congress or either state legislative chamber. 

However, imagine an alternative rule in which a commissioner or special master was told to choose from among the relevant neutral ensemble a plan for which the anticipated seat share of each party was 50 percent when the vote share was 50 percent.  At the scale of Congressional districts or state senate districts, the range of outcomes in the ensemble is sufficiently large that this could be achieved by selecting one of the most pro-Democratic plans.  However, this becomes impossible as districts become smaller and more numerous.  The range of outcomes is much narrower at the scale of Pennsylvania House districts, where even the most Democratic plan falls short of proportionality.  To be clear, the lesson is not that a ``fair'' plan with 203 districts cannot be drawn in Pennsylvania.  Rather, such a plan does not emerge from the neutral ensembles, and it might take a conscious effort to consider partisanship in order to produce one.                

Some interesting inferences---and questions for further analysis---emerge from comparisons of the graphs for different elections.  One lesson, explored further below, is that the statewide vote share is important.  The Democrats' seat share is especially far from proportionality when their vote share is low (e.g. GOV10), and they are still quite far from proportionality even in elections that are very close to 50 percent.  In fact, even in an election with 55 percent of the vote (SEN12), they do not achieve proportionality.  Only when they received 57 percent of the vote (ATG12) did they significantly surpass proportionality.  

This latter comparison suggests that perhaps there are differences between these two races that go beyond the difference in vote shares between SEN12 and ATG12.  In the Attorney General election, the Democratic candidate, Kathleen Kane, outperformed the Democratic Senate candidate, Bob Casey, who was on the same ballot on the same day, by 2.86 percentage points.  But the difference in seats was substantially larger.  At the scale of Congressional districts and state senate seats, Casey came out ahead, on average, in less than 55 percent of the districts, while Kane came out ahead in well over 65 percent.  This indicates that it matters not only that Kane received more votes than Casey, but also \emph{where} she outperformed him.  That is to say, she had stronger support than Casey in some geographic areas where, in the ensembles, Casey fell below 50 percent.  For instance, Kane outperformed Casey in many of the counties surrounding her home town of Scranton, as well as in the counties along the Western border of the state.        

Another election pair that stands out as a place where subtle geographic factors play a big role is PRES16 and SEN16. These two elections were on the same ballot and their statewide shares differed by a mere 0.37 percentage points, yet the Democrats' ability to turn votes into seats in a neutral redistricting process is substantially worse in SEN16 than in PRES16.  In other words, as with Kathleen Kane vis-a-vis Bob Casey, Hillary Clinton was stronger than the Democratic Senate candidate, Katie McGinty, in parts of the state that leaned Republican---as it turns out, parts of suburban Philadelphia---in the Senate race.  

We should take this as a warning that subtle changes in voting patterns can result in significant swings in representation that elude simple explanation. It is true that in the era of polarization and nationalized politics, results of various statewide races are highly correlated.  Nevertheless, split-ticket voting is still alive and well, and the spatial distribution of votes varies across races in ways that are consequential for inferences about representation.   

\section{Seats-votes plots}
In the previous section, we broke down the data by election. In this section, we will plot all elections together for each of three districting scales. We organize the elections by their statewide vote-share to produce a seats-votes plot.  The values on the horizontal axis correspond to the observed statewide vote share in each of the nine statewide elections examined above. This is meant to parallel the traditional seats-votes curves used in partisan symmetry analysis (see e.g.~\cite{grofman83}). However, our plots contain far more information than curves since they cover the full range of possibilities encountered in a neutral ensemble associated with each election.

\begin{figure}
\centering
\foreach \k in {18, 50, 203}
{
\begin{subfigure}{0.46\textwidth}
\centering
\resizebox{\columnwidth}{!}{%
\begin{tikzpicture}
\begin{axis}[
xmin=0, xmax=1, ymin=0, ymax=1,
colormap/jet, mark size=1,
 axis equal image,
 title=\k\ districts
]
\addplot+[scatter, only marks, mark=*, scatter src=explicit] table[x=X, y=Y, col sep=comma, meta=s, forget plot]{RoddenWeighill/seatsvotes_\k .csv};
\addplot[gray, forget plot] coordinates{(0,0) (1,1)};
\addplot[green, forget plot] coordinates{(0.25,0) (0.75,1)};
\end{axis}
\end{tikzpicture}
}
\end{subfigure}

}
\caption{Seats-votes plots for Pennsylvania. The $x$-axis indicates statewide Democratic vote share and the $y$-axis indicates Democratic seat share in each case. Grey lines indicate proportionality, while green lines indicate an efficiency gap of zero.}
\label{seatsvotes}
\end{figure}
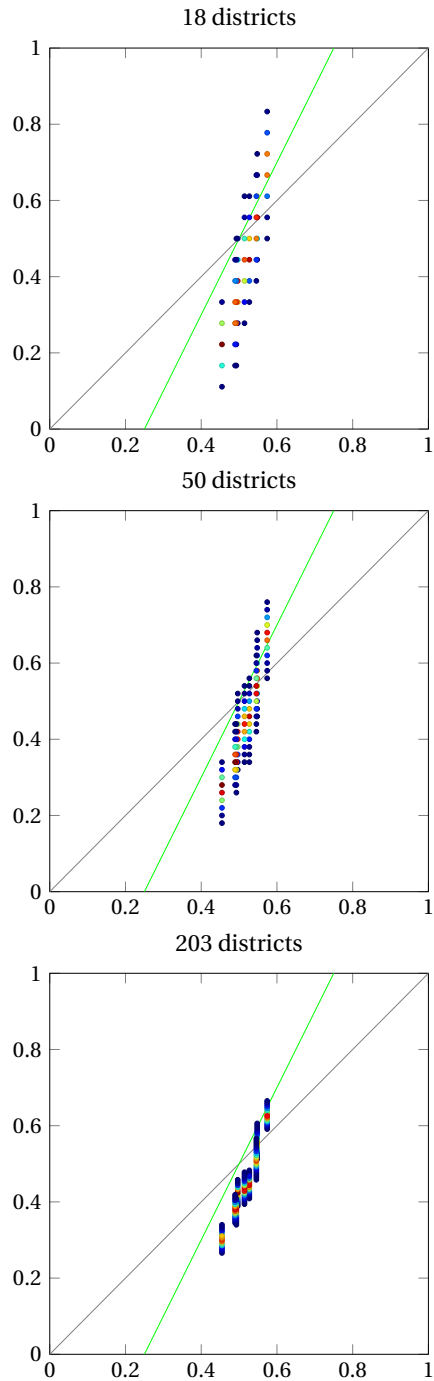

In each of the figures in Figure \ref{seatsvotes}, we have selected points from the plots in the previous section which correspond to a particular scale of redistricting. Each dot therefore represents a set of districting plans. The colors are the same; lighter colors represent more frequent seat share outcomes. What has changed is the $x$-axis, which now represents the statewide vote share of the election used to calculate the seat share. The dotted lines indicate two different doctrines of ``fairness'' one might adopt which are not based on ensembles. The gray dotted line indicates proportionality (that is, seat share equals vote share). The green line indicates outcomes which correspond to an efficiency gap of zero.\footnote{This is the simplified EG formula, assuming equal turnout in each district.} Efficiency gap is a measure of fairness found in the literature based on the concept of ``wasted votes'' \cite{stephanopoulos2015partisan}.

The Republican advantage we observed in the previous section is strikingly visible in these plots as well. Where the dots are below the gray line, Democrats are under-performing relative to proportionality: this is the case for all but the two most Democratic elections considered (ATG12 and SEN12).  A good way to appreciate the asymmetry present in these graphs is to contrast elections in which Democrats receive around 55 percent of the vote (SEN12 and GOV14) with an election in which the Republican candidate received around 55 percent of the vote (GOV10).  At the scale of Congressional districts, on average, the neutral plans produce an expected seat share of around 54 percent for Democrats, but around 77 percent for Republicans.  

While proportionality is considered by some as the mark of a fair districting process, others recognize that a ``winner's bonus'' is a reasonable property to expect in a districted system. That is, while 50\% of the vote should win you half the seats, 70\% of the vote---an overwhelming victory---could easily result in far more than 70\% of the seats depending on how the extra 20\% advantage is spatially arranged. If it is uniform, of course, then all seats go to the majority party.  The efficiency gap boils down to a specific recommendation for this bonus: $50 + x$ percent of the vote should roughly translate into $50 + 2x$ of the seats. This is where the green line comes from. All this discussion is to say that where the dots are below the green line for Democratic vote shares less than $0.5$, the Democrats are not only failing to achieve proportionality, but are not even able to achieve the representation predicted by a common standard in the literature which takes into account the Republican winner's bonus.

An interesting but subtle scale phenomenon is visible on these plots for the most Democratic election of all---ATG12. With 57 percent of the statewide vote, Democrats exceed the efficiency gap standard for 18 and 50 districts for many plans, but far more rarely for the plans with 203 districts. In other words, the Democratic winner's bonus diminishes as the scale of redistricting grows finer for this particular election.  

And again, these graphs demonstrate the much tighter range of outcomes produced in the neutral ensembles as districts become smaller.  As mentioned above, some reformers anticipate that smaller districts on the scale of Canadian or British parliamentary constituencies, or Pennsylvania State House districts, might reduce the level of Republican advantage observed in recent Congressional elections.  However, these graphs suggest that in Pennsylvania, neutral districting at a smaller scale might not produce any maps at all that meet a rather uncontroversial standard of partisan fairness.   

\section{East versus West}

\begin{center}
    \includegraphics[width=\textwidth]{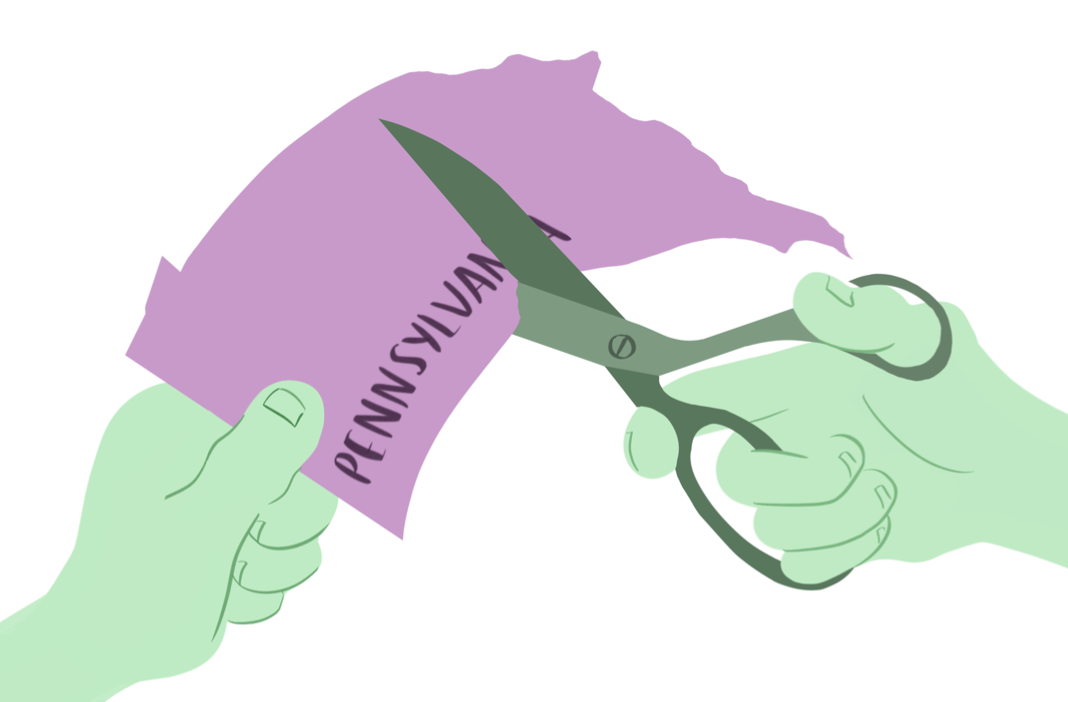}
\end{center}

\subsection*{Method}
\begin{figure}
\centering
\includegraphics[width=0.6\textwidth]{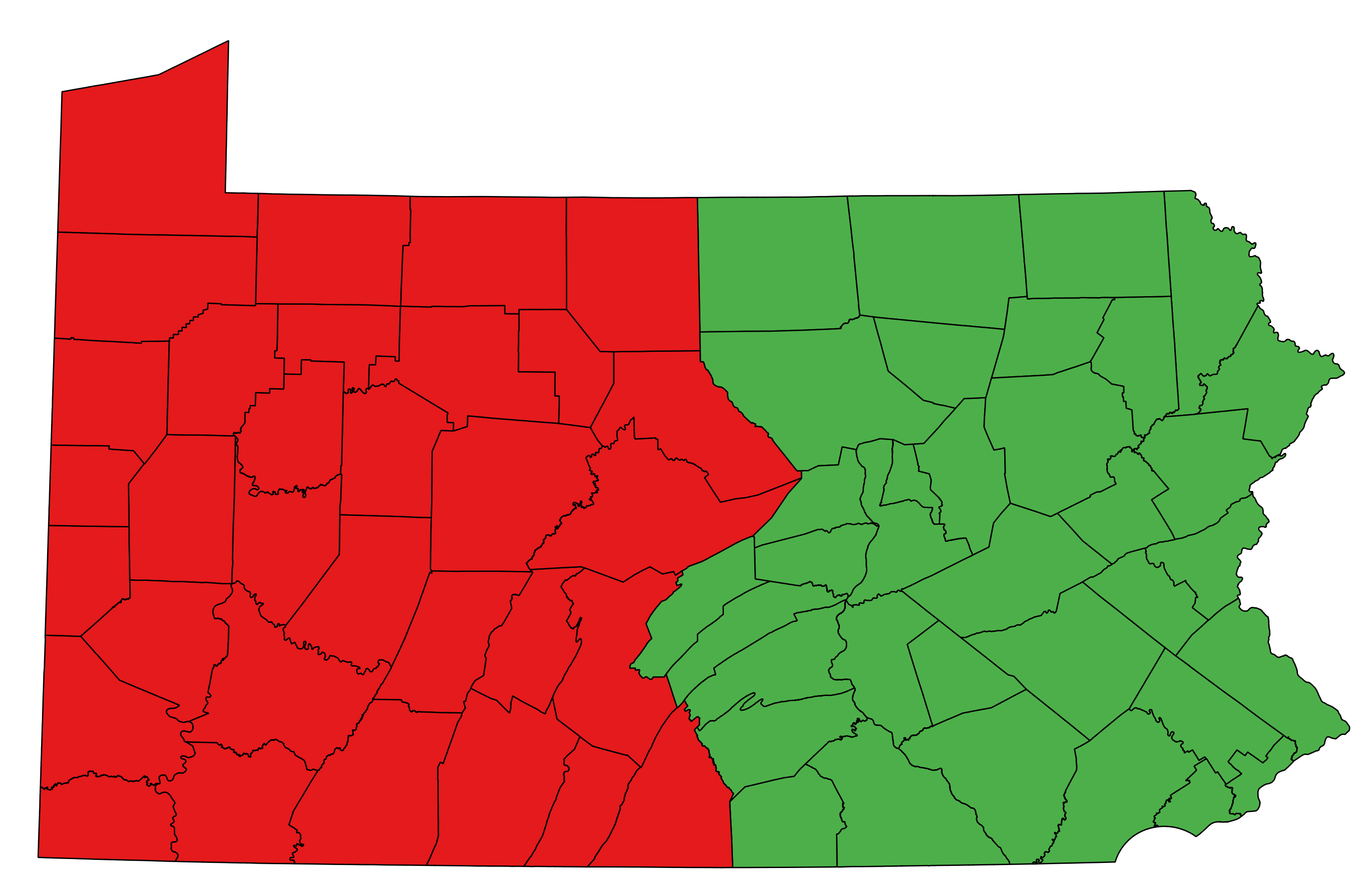}
\caption{Dividing Pennsylvania into West and East.}
\label{divide}
\end{figure}

In this section we examine the difference in political geography between the western and eastern parts of Pennsylvania at the level of Congressional and Pennsylvania state Senate redistricting (we omit the state House level for the sake of brevity). We choose a subdivision of Pennsylvania along county boundaries which results in two pieces, West and East, as shown in Figure \ref{divide}. Up to an error of just over 3000 people, the West has half the population of the East. Since Pennsylvania has 18 congressional districts, it thus makes sense to consider plans of six districts for the West and plans of twelve districts for the East. To best approximate state Senate plans, we consider plans of seventeen districts for the West and plans of thirty-four districts for the East, adding up to a total of fifty-one, one over the correct number of fifty. For ease of reference, the overall vote shares for each party in each piece are shown in Tables \ref{statewidetable} and \ref{statewidetable51}, along with the mean seat shares coming from the ensemble analyses.

\subsection*{Results}

The Figures in this section each have four histograms showing the Democratic seats outcomes for four different ensembles based on the specified election data. The ``West'' ensemble is an ensemble of 50,000 plans with a third of the targeted number of districts for the West piece of Pennsylvania only. The ``East'' ensemble is an ensemble of 50,000 plans with two-thirds of the targeted number of districts for the East piece. The ``Full state'' ensemble is an ensemble of 50,000 plans for the entire state with the targeted number of districts. Finally, the ``E-W pairs'' ensemble consists of every possible plan that can be created by putting a plan from the ``West'' ensemble and a plan from the ``East'' ensemble put together. This last ensemble should be thought of as an ensemble of plans that respect the West-East subdivision of the state we chose. As mentioned above, we chose two districting scales: 18 districts (for Congressional) and 51 districts (as the closest multiple of three to the state Senate number of 50).

\begin{table}
\centering
\begin{tabular}{|c||c|c||c|c||c|c|}
\hline
& \multicolumn{2}{c||}{West} & \multicolumn{2}{c||}{East} & \multicolumn{2}{c|}{Full}  \\ \hline
& seat \%  & vote \%  & seat \% & vote \% & seat \% & vote \% \\ \hline
PRES16  & 20.76\% & 41.66\% & 46.83\% & 53.60\% & 37.83\% & 49.65\% \\
\hline
SEN16  & 22.08\% & 43.36\% & 36.32\% & 52.18\% & 31.62\% & 49.28\% \\
\hline
ATG16  & 27.22\% & 46.04\% & 51.18\% & 54.11\% & 43.61\% & 51.43\% \\
\hline
PRES12  & 23.49\% & 46.03\% & 57.47\% & 56.04\% & 46.46\% & 52.71\% \\
\hline
SEN12  & 33.64\% & 48.83\% & 62.79\% & 57.45\% & 53.05\% & 54.56\% \\
\hline
ATG12  & 67.55\% & 54.00\% & 69.74\% & 59.12\% & 69.23\% & 57.42\% \\
\hline
GOV14  & 38.57\% & 49.62\% & 64.54\% & 57.30\% & 56.27\% & 54.78\% \\
\hline
GOV10  & 10.59\% & 40.49\% & 29.07\% & 48.08\% & 22.69\% & 45.48\% \\
\hline
SEN10  & 20.72\% & 45.14\% & 36.26\% & 50.92\% & 31.23\% & 48.95\% \\
\hline
\end{tabular}
\caption{Vote shares and mean seat shares for 18 districts (6 West, 12 East).}
\label{statewidetable}
\end{table}

\begin{table}
\centering
\begin{tabular}{|c||c|c||c|c||c|c|}
\hline
& \multicolumn{2}{c||}{West} & \multicolumn{2}{c||}{East} & \multicolumn{2}{c|}{Full}  \\ \hline
& seat \%  & vote \%  & seat \% & vote \% & seat \% & vote \% \\ \hline
PRES16  & 24.85\% & 41.66\% & 49.64\% & 53.60\% & 41.90\% & 49.65\% \\
\hline
SEN16  & 20.59\% & 43.36\% & 42.40\% & 52.18\% & 35.63\% & 49.28\% \\
\hline
ATG16  & 28.60\% & 46.04\% & 51.30\% & 54.11\% & 43.96\% & 51.43\% \\
\hline
PRES12  & 25.70\% & 46.03\% & 55.74\% & 56.04\% & 46.23\% & 52.71\% \\
\hline
SEN12  & 35.02\% & 48.83\% & 61.57\% & 57.45\% & 52.91\% & 54.56\% \\
\hline
ATG12  & 63.25\% & 54.00\% & 69.84\% & 59.12\% & 67.30\% & 57.42\% \\
\hline
GOV14  & 41.90\% & 49.62\% & 65.23\% & 57.30\% & 56.77\% & 54.78\% \\
\hline
GOV10  & 17.78\% & 40.49\% & 32.03\% & 48.08\% & 27.16\% & 45.48\% \\
\hline
SEN10  & 26.00\% & 45.14\% & 38.89\% & 50.92\% & 34.96\% & 48.95\% \\
\hline
\end{tabular}
\caption{Vote shares and mean seat shares for 51 districts (17 West, 34 East).}
\label{statewidetable51}
\end{table}

%%%%%E-W HISTOGRAMS%%%%%%%
\begin{figure}
\foreach \e in {PRES16, SEN16, ATG16}
{
\begin{subfigure}{\textwidth}
\caption*{\underline{\e}}

\vspace{-0.3cm}

\centering

\begin{tikzpicture}
\begin{axis}[grid=none, xtick distance=1,
ymax=1, ymin=0, clip=false,
title=West (6),
width=3.75cm,
height=3.75cm,
tick label style={font=\tiny},
label style={font=\tiny},
title style={font=\footnotesize,yshift=-1ex},
at={(0cm,3cm)}
]
\addplot[green, dashed, line width=1pt] table[x=W, y=Y, col sep=comma]{RoddenWeighill/voteshares\e _18.csv};
\addplot[black, fill=red, ybar interval] table[x=WX, y=WY, col sep=comma]{RoddenWeighill/Whist\e .csv};
\end{axis}

\begin{axis}[grid=none, xtick distance=1,
ymax=1, ymin=0, clip=false,
title=East (12),
width=3.75cm,
height=3.75cm,
yticklabels={,,},
tick label style={font=\tiny},
label style={font=\tiny} ,
title style={font=\footnotesize,yshift=-1ex},
at={(3cm,3cm)}
]
\addplot[green, dashed, line width=1pt] table[x=E, y=Y, col sep=comma]{RoddenWeighill/voteshares\e _18.csv};
\addplot[ybar interval, black, fill= green!60!black] table[x=EX, y=EY, col sep=comma]{RoddenWeighill/Ehist\e .csv};
\end{axis}

\begin{axis}[grid=none, xtick distance=2,
ymax=1, ymin=0, clip=false,
title=Full state (18),
width=3.75cm,
height=3.75cm,
yticklabels={,,},
tick label style={font=\tiny} ,
label style={font=\tiny},
title style={font=\footnotesize,yshift=-1ex},
at={(6cm,3cm)}
]
\addplot[green, dashed, line width=1pt] table[x=F, y=Y, col sep=comma]{RoddenWeighill/voteshares\e _18.csv};
\addplot[ybar interval, black, fill=purple] table[x=FX, y=FY, col sep=comma]{RoddenWeighill/Fhist\e .csv};
\end{axis}

\begin{axis}[grid=none, xtick distance=2,
ymax=1, ymin=0, clip=false,
title=E-W pairs (18),
width=3.75cm,
height=3.75cm,
yticklabels={,,},
tick label style={font=\tiny} ,
label style={font=\tiny},
title style={font=\footnotesize,yshift=-1ex},
at={(9cm,3cm)}
]
\addplot[green, dashed, line width=1pt] table[x=B, y=Y, col sep=comma]{RoddenWeighill/voteshares\e _18.csv};
\addplot[ybar interval, black, fill=brown] table[x=BX, y=BY, col sep=comma]{RoddenWeighill/Bhist\e .csv};
\end{axis}

\begin{axis}[grid=none, xtick distance=1,
ymax=1, ymin=0, clip=false,
title=West (17),
width=3.75cm,
height=3.75cm,
tick label style={font=\tiny},
title style={font=\footnotesize,yshift=-1ex},
at={(0cm,0cm)}
]
\addplot[green, dashed, line width=1pt] table[x=W, y=Y, col sep=comma]{RoddenWeighill/voteshares\e _51.csv};

\addplot[black, fill=red, ybar interval] table[x=WX, y=WY, col sep=comma]{RoddenWeighill/Whist\e _51.csv};
\end{axis}

\begin{axis}[grid=none, xtick distance=2,
ymax=1, ymin=0, clip=false,
title=East (34),
width=3.75cm,
height=3.75cm,
yticklabels={,,},
tick label style={font=\tiny},
title style={font=\footnotesize,yshift=-1ex},
at={(3cm,0cm)}
]
\addplot[green, dashed, line width=1pt] table[x=E, y=Y, col sep=comma]{RoddenWeighill/voteshares\e _51.csv};

\addplot[ybar interval, black, fill= green!60!black] table[x=EX, y=EY, col sep=comma]{RoddenWeighill/Ehist\e _51.csv};
\end{axis}

\begin{axis}[grid=none, xtick distance=3,
ymax=1, ymin=0, clip=false,
title=Full state (51),
width=3.75cm,
height=3.75cm,
yticklabels={,,},
tick label style={font=\tiny},
title style={font=\footnotesize,yshift=-1ex},
at={(6cm,0cm)}
]
\addplot[green, dashed, line width=1pt] table[x=F, y=Y, col sep=comma]{RoddenWeighill/voteshares\e _51.csv};
\addplot[ybar interval, black, fill=purple] table[x=FX, y=FY, col sep=comma]{RoddenWeighill/Fhist\e _51.csv};
\end{axis}

\begin{axis}[grid=none, xtick distance=3,
ymax=1, ymin=0, clip=false,
title=E-W pairs (51),
width=3.75cm,
height=3.75cm,
yticklabels={,,},
tick label style={font=\tiny},
title style={font=\footnotesize,yshift=-1ex},
at={(9cm,0cm)}
]
\addplot[green, dashed, line width=1pt] table[x=B, y=Y, col sep=comma]{RoddenWeighill/voteshares\e _51.csv};

\addplot[ybar interval, black, fill=brown] table[x=BX, y=BY, col sep=comma]{RoddenWeighill/Bhist\e _51.csv};
\end{axis}
\end{tikzpicture}
\end{subfigure}
}
\caption{East-West comparison for 2016 elections. The $x$-axis and $y$-axis in each plot represent Democratic seats won and fraction of the ensemble respectively. Numbers in parentheses indicate the number of districts in each plan. Dotted green lines indicate proportionality.}
\end{figure}
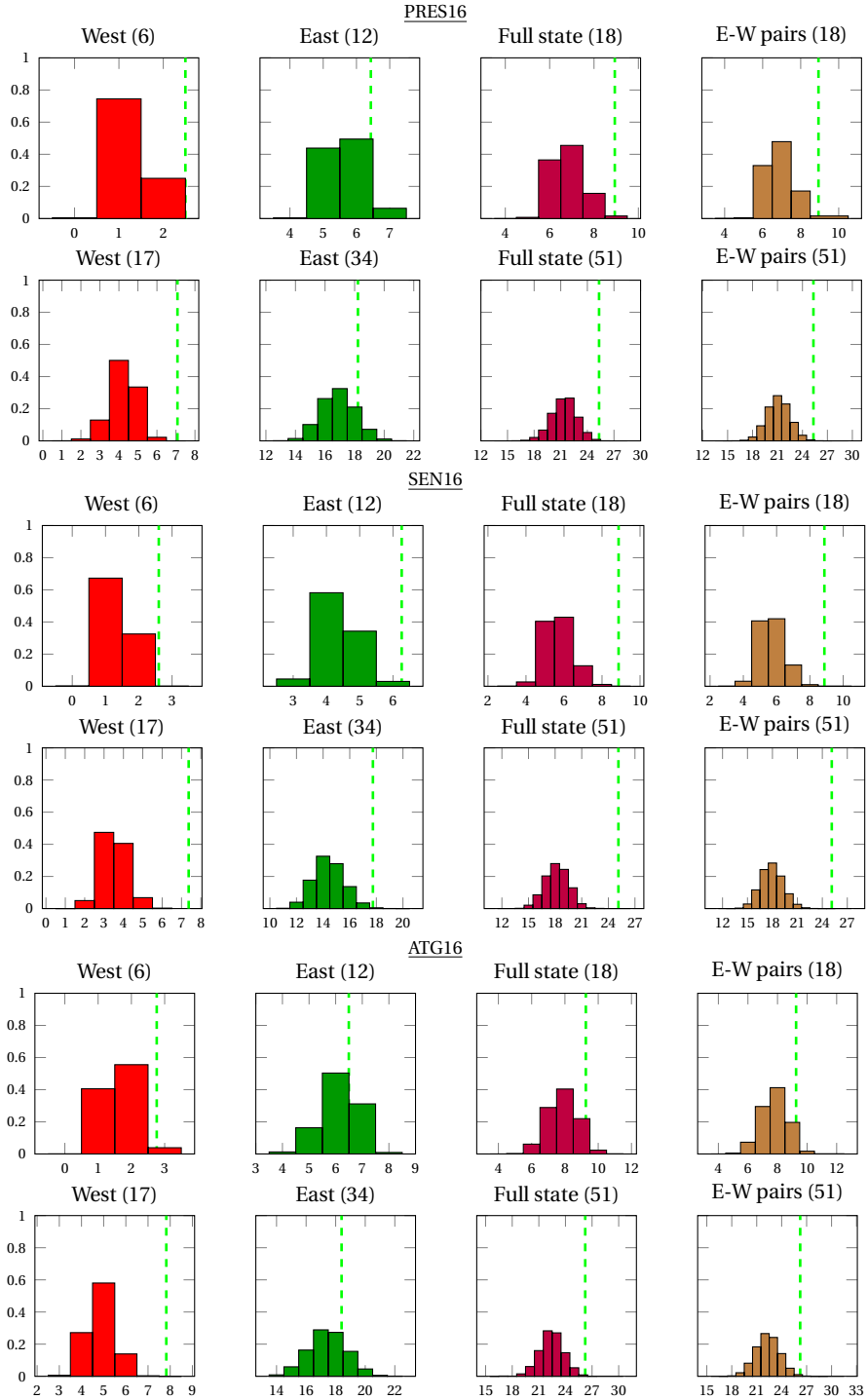

%%%%%%%%%%%%%%
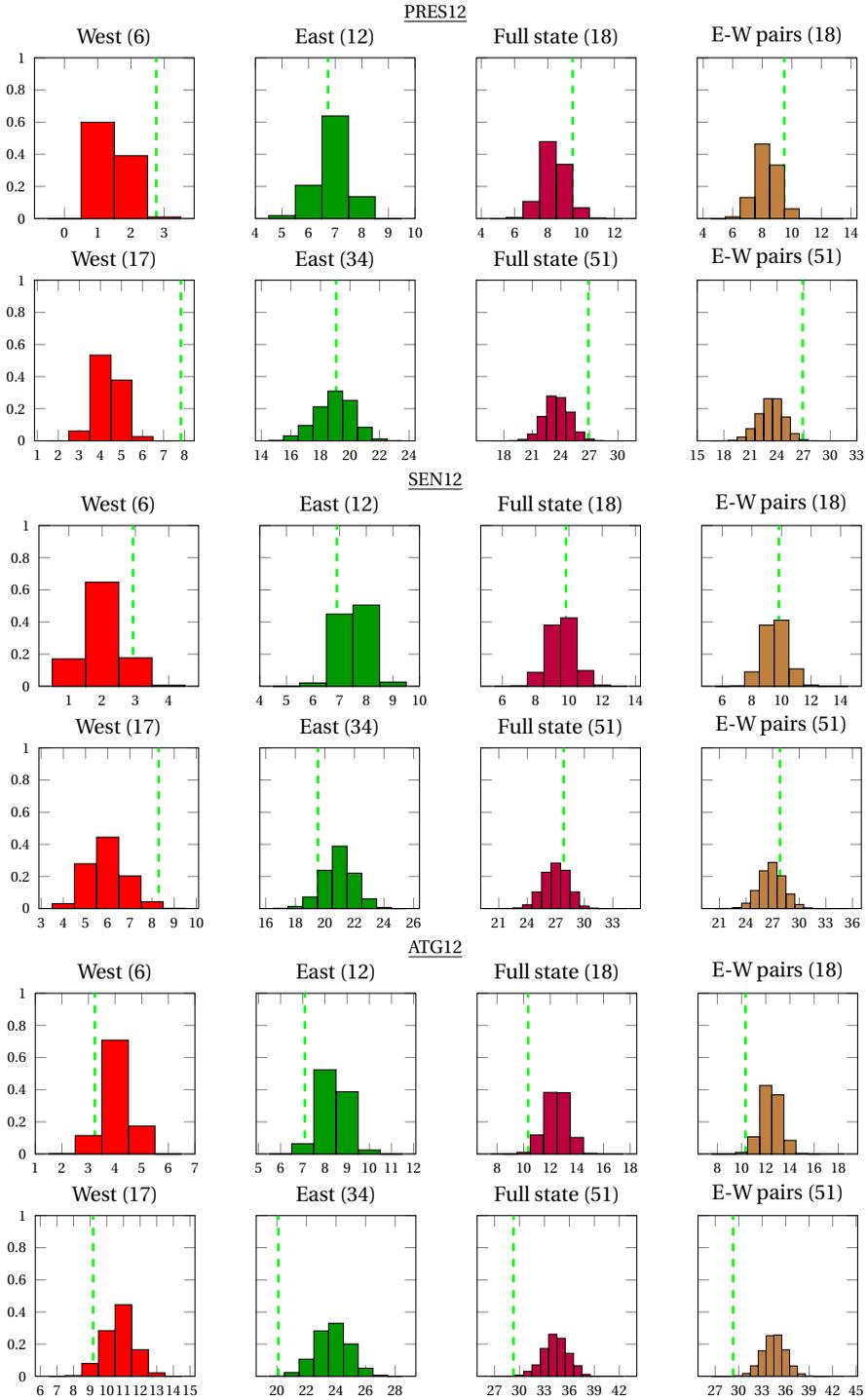
\begin{figure}
\foreach \e in {PRES12, SEN12, ATG12}
{
\begin{subfigure}{\textwidth}
\caption*{\underline{\e}}

\vspace{-0.3cm}

\centering

\begin{tikzpicture}
\begin{axis}[grid=none, xtick distance=1,
ymax=1, ymin=0, clip=false,
title=West (6),
width=3.75cm,
height=3.75cm,
tick label style={font=\tiny},
label style={font=\tiny},
title style={font=\footnotesize,yshift=-1ex},
at={(0cm,3cm)}
]
\addplot[green, dashed, line width=1pt] table[x=W, y=Y, col sep=comma]{RoddenWeighill/voteshares\e _18.csv};
\addplot[black, fill=red, ybar interval] table[x=WX, y=WY, col sep=comma]{RoddenWeighill/Whist\e .csv};
\end{axis}

\begin{axis}[grid=none, xtick distance=1,
ymax=1, ymin=0, clip=false,
title=East (12),
width=3.75cm,
height=3.75cm,
yticklabels={,,},
tick label style={font=\tiny},
label style={font=\tiny} ,
title style={font=\footnotesize,yshift=-1ex},
at={(3cm,3cm)}
]
\addplot[green, dashed, line width=1pt] table[x=E, y=Y, col sep=comma]{RoddenWeighill/voteshares\e _18.csv};
\addplot[ybar interval, black, fill= green!60!black] table[x=EX, y=EY, col sep=comma]{RoddenWeighill/Ehist\e .csv};
\end{axis}

\begin{axis}[grid=none, xtick distance=2,
ymax=1, ymin=0, clip=false,
title=Full state (18),
width=3.75cm,
height=3.75cm,
yticklabels={,,},
tick label style={font=\tiny} ,
label style={font=\tiny},
title style={font=\footnotesize,yshift=-1ex},
at={(6cm,3cm)}
]
\addplot[green, dashed, line width=1pt] table[x=F, y=Y, col sep=comma]{RoddenWeighill/voteshares\e _18.csv};
\addplot[ybar interval, black, fill=purple] table[x=FX, y=FY, col sep=comma]{RoddenWeighill/Fhist\e .csv};
\end{axis}

\begin{axis}[grid=none, xtick distance=2,
ymax=1, ymin=0, clip=false,
title=E-W pairs (18),
width=3.75cm,
height=3.75cm,
yticklabels={,,},
tick label style={font=\tiny} ,
label style={font=\tiny},
title style={font=\footnotesize,yshift=-1ex},
at={(9cm,3cm)}
]
\addplot[green, dashed, line width=1pt] table[x=B, y=Y, col sep=comma]{RoddenWeighill/voteshares\e _18.csv};
\addplot[ybar interval, black, fill=brown] table[x=BX, y=BY, col sep=comma]{RoddenWeighill/Bhist\e .csv};
\end{axis}

\begin{axis}[grid=none, xtick distance=1,
ymax=1, ymin=0, clip=false,
title=West (17),
width=3.75cm,
height=3.75cm,
tick label style={font=\tiny},
title style={font=\footnotesize,yshift=-1ex},
at={(0cm,0cm)}
]
\addplot[green, dashed, line width=1pt] table[x=W, y=Y, col sep=comma]{RoddenWeighill/voteshares\e _51.csv};

\addplot[black, fill=red, ybar interval] table[x=WX, y=WY, col sep=comma]{RoddenWeighill/Whist\e _51.csv};
\end{axis}

\begin{axis}[grid=none, xtick distance=2,
ymax=1, ymin=0, clip=false,
title=East (34),
width=3.75cm,
height=3.75cm,
yticklabels={,,},
tick label style={font=\tiny},
title style={font=\footnotesize,yshift=-1ex},
at={(3cm,0cm)}
]
\addplot[green, dashed, line width=1pt] table[x=E, y=Y, col sep=comma]{RoddenWeighill/voteshares\e _51.csv};

\addplot[ybar interval, black, fill= green!60!black] table[x=EX, y=EY, col sep=comma]{RoddenWeighill/Ehist\e _51.csv};
\end{axis}

\begin{axis}[grid=none, xtick distance=3,
ymax=1, ymin=0, clip=false,
title=Full state (51),
width=3.75cm,
height=3.75cm,
yticklabels={,,},
tick label style={font=\tiny},
title style={font=\footnotesize,yshift=-1ex},
at={(6cm,0cm)}
]
\addplot[green, dashed, line width=1pt] table[x=F, y=Y, col sep=comma]{RoddenWeighill/voteshares\e _51.csv};
\addplot[ybar interval, black, fill=purple] table[x=FX, y=FY, col sep=comma]{RoddenWeighill/Fhist\e _51.csv};
\end{axis}

\begin{axis}[grid=none, xtick distance=3,
ymax=1, ymin=0, clip=false,
title=E-W pairs (51),
width=3.75cm,
height=3.75cm,
yticklabels={,,},
tick label style={font=\tiny},
title style={font=\footnotesize,yshift=-1ex},
at={(9cm,0cm)}
]
\addplot[green, dashed, line width=1pt] table[x=B, y=Y, col sep=comma]{RoddenWeighill/voteshares\e _51.csv};

\addplot[ybar interval, black, fill=brown] table[x=BX, y=BY, col sep=comma]{RoddenWeighill/Bhist\e _51.csv};
\end{axis}
\end{tikzpicture}
\end{subfigure}
}
\caption{East-West comparison for 2012 elections. The $x$-axis and $y$-axis in each plot represent Democratic seats won and fraction of the ensemble respectively. Numbers in parentheses indicate the number of districts in each plan. Dotted green lines indicate proportionality.}
\end{figure}

%%%%%%%%%%%%%%%%%%%
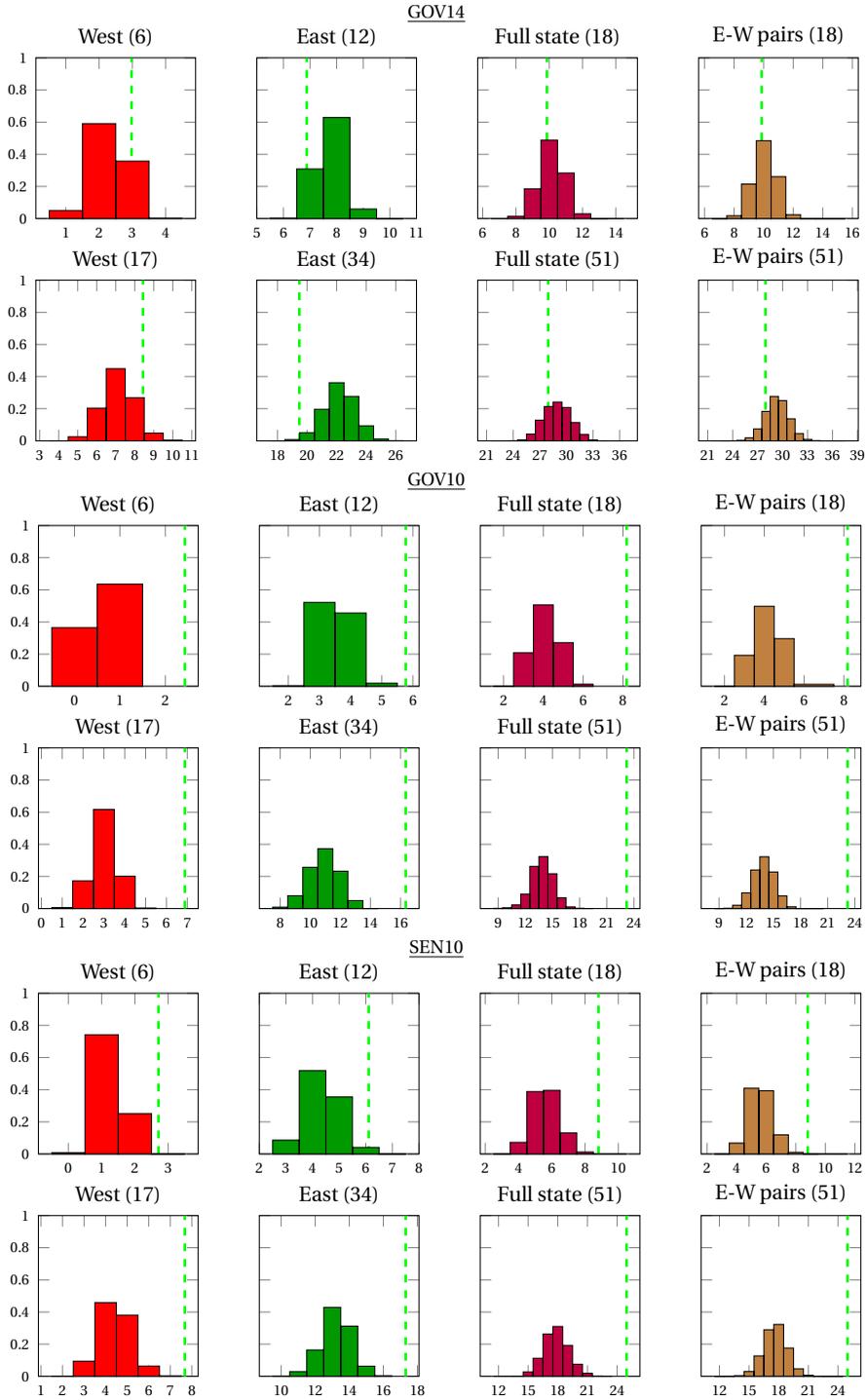
\begin{figure}
\foreach \e in {GOV14, GOV10, SEN10}
{
\begin{subfigure}{\textwidth}
\caption*{\underline{\e}}

\vspace{-0.3cm}

\centering

\begin{tikzpicture}
\begin{axis}[grid=none, xtick distance=1,
ymax=1, ymin=0, clip=false,
title=West (6),
width=3.75cm,
height=3.75cm,
tick label style={font=\tiny},
label style={font=\tiny},
title style={font=\footnotesize,yshift=-1ex},
at={(0cm,3cm)}
]
\addplot[green, dashed, line width=1pt] table[x=W, y=Y, col sep=comma]{RoddenWeighill/voteshares\e _18.csv};
\addplot[black, fill=red, ybar interval] table[x=WX, y=WY, col sep=comma]{RoddenWeighill/Whist\e .csv};
\end{axis}

\begin{axis}[grid=none, xtick distance=1,
ymax=1, ymin=0, clip=false,
title=East (12),
width=3.75cm,
height=3.75cm,
yticklabels={,,},
tick label style={font=\tiny},
label style={font=\tiny} ,
title style={font=\footnotesize,yshift=-1ex},
at={(3cm,3cm)}
]
\addplot[green, dashed, line width=1pt] table[x=E, y=Y, col sep=comma]{RoddenWeighill/voteshares\e _18.csv};
\addplot[ybar interval, black, fill= green!60!black] table[x=EX, y=EY, col sep=comma]{RoddenWeighill/Ehist\e .csv};
\end{axis}

\begin{axis}[grid=none, xtick distance=2,
ymax=1, ymin=0, clip=false,
title=Full state (18),
width=3.75cm,
height=3.75cm,
yticklabels={,,},
tick label style={font=\tiny} ,
label style={font=\tiny},
title style={font=\footnotesize,yshift=-1ex},
at={(6cm,3cm)}
]
\addplot[green, dashed, line width=1pt] table[x=F, y=Y, col sep=comma]{RoddenWeighill/voteshares\e _18.csv};
\addplot[ybar interval, black, fill=purple] table[x=FX, y=FY, col sep=comma]{RoddenWeighill/Fhist\e .csv};
\end{axis}

\begin{axis}[grid=none, xtick distance=2,
ymax=1, ymin=0, clip=false,
title=E-W pairs (18),
width=3.75cm,
height=3.75cm,
yticklabels={,,},
tick label style={font=\tiny} ,
label style={font=\tiny},
title style={font=\footnotesize,yshift=-1ex},
at={(9cm,3cm)}
]
\addplot[green, dashed, line width=1pt] table[x=B, y=Y, col sep=comma]{RoddenWeighill/voteshares\e _18.csv};
\addplot[ybar interval, black, fill=brown] table[x=BX, y=BY, col sep=comma]{RoddenWeighill/Bhist\e .csv};
\end{axis}

\begin{axis}[grid=none, xtick distance=1,
ymax=1, ymin=0, clip=false,
title=West (17),
width=3.75cm,
height=3.75cm,
tick label style={font=\tiny},
title style={font=\footnotesize,yshift=-1ex},
at={(0cm,0cm)}
]
\addplot[green, dashed, line width=1pt] table[x=W, y=Y, col sep=comma]{RoddenWeighill/voteshares\e _51.csv};

\addplot[black, fill=red, ybar interval] table[x=WX, y=WY, col sep=comma]{RoddenWeighill/Whist\e _51.csv};
\end{axis}

\begin{axis}[grid=none, xtick distance=2,
ymax=1, ymin=0, clip=false,
title=East (34),
width=3.75cm,
height=3.75cm,
yticklabels={,,},
tick label style={font=\tiny},
title style={font=\footnotesize,yshift=-1ex},
at={(3cm,0cm)}
]
\addplot[green, dashed, line width=1pt] table[x=E, y=Y, col sep=comma]{RoddenWeighill/voteshares\e _51.csv};

\addplot[ybar interval, black, fill= green!60!black] table[x=EX, y=EY, col sep=comma]{RoddenWeighill/Ehist\e _51.csv};
\end{axis}

\begin{axis}[grid=none, xtick distance=3,
ymax=1, ymin=0, clip=false,
title=Full state (51),
width=3.75cm,
height=3.75cm,
yticklabels={,,},
tick label style={font=\tiny},
title style={font=\footnotesize,yshift=-1ex},
at={(6cm,0cm)}
]
\addplot[green, dashed, line width=1pt] table[x=F, y=Y, col sep=comma]{RoddenWeighill/voteshares\e _51.csv};
\addplot[ybar interval, black, fill=purple] table[x=FX, y=FY, col sep=comma]{RoddenWeighill/Fhist\e _51.csv};
\end{axis}

\begin{axis}[grid=none, xtick distance=3,
ymax=1, ymin=0, clip=false,
title=E-W pairs (51),
width=3.75cm,
height=3.75cm,
yticklabels={,,},
tick label style={font=\tiny},
title style={font=\footnotesize,yshift=-1ex},
at={(9cm,0cm)}
]
\addplot[green, dashed, line width=1pt] table[x=B, y=Y, col sep=comma]{RoddenWeighill/voteshares\e _51.csv};

\addplot[ybar interval, black, fill=brown] table[x=BX, y=BY, col sep=comma]{RoddenWeighill/Bhist\e _51.csv};
\end{axis}
\end{tikzpicture}
\end{subfigure}
}
\caption{East-West comparison for other elections. The $x$-axis and $y$-axis in each plot represent Democratic seats won and fraction of the ensemble respectively. Numbers in parentheses indicate the number of districts in each plan. Dotted green lines indicate proportionality.}
\end{figure}

\subsection*{Discussion}
One observation we should immediately make is that the seats outcomes for the unsplit state plans and the East-West combination plans are in all cases extremely similar. In other words, forcing plans to respect our arbitrary East-West division does not have a substantial impact on the baseline for redistricting in Pennsylvania. This gives us the confidence to examine the impacts of the East and West on baseline representation separately, since combining them pairwise reproduces the redistricting phenomena we are trying to study for the whole state.

The plots reveal that the general Democratic under-performance is more pronounced in the West than in the East. In the West, in both PRES16 and SEN16, the Democrats were able to secure only one Western Congressional seat in a majority of the plans (in Pittsburgh), despite the Western vote share being well above 40\% in both cases.  Even when the Democrats receive 49 percent of the votes in the West, as they did in SEN12, they only received 34 percent of the Congressional seats.   There is some contrast between the two elections where the Democrats achieve a higher mean seat share than vote share. For ATG12, when Kane received a statewide vote share of 57 percent, both the West and East mean seat share exceed the vote share (the West by a greater margin than the East in fact). For GOV14, when the Democratic candidate received 55 percent statewide, the mean seat share falls short of the vote share in the West but not the East, and the two combine to result in a statewide seat share that is slightly above the statewide vote share.

The political geography of Western Pennsylvania seems to make it quite difficult for the Democrats to transform votes to seats.  At the scale of Congressional districts, in a typical election, the ensembles tend to produce a single Democratic Pittsburgh seat.  Perhaps there is a hint of a scale effect here, since the Democratic seat share is somewhat higher at the scale of state Senate than Congressional districts in the West for 7 of the 9 elections.  This may have to do with the nature of the partitioning of Pittsburgh, and the greater likelihood of Democratic victories occurring in Erie at the smaller scale of state Senate districts.  To be sure, the Democrats' political geography is still quite inefficient in the East, but the Democrats' difficulty in the West is especially striking at both spatial scales analyzed here.  

The East-West comparison is also useful for shedding light on the puzzling gap, described above, between SEN16 and PRES16.  The right-hand columns of Tables 1 and 2 illuminate that in the state as a whole, Clinton's presidential vote share was more efficiently distributed than that of McGinty in the Senate race.  With very similar vote shares, on average, neutral ensembles produced a seat share about 6 percentage points higher for Clinton at both spatial scales considered here.  We can now see that McGinty outperformed Clinton in the West, and Clinton outperformed McGinty in the East.  Inspection of precinct-level maps reveals that split-ticket voters favoring the Democratic Senate candidate while favoring Donald Trump in the presidential election in the West were located in non-urban working-class areas, especially in the Southwest.  And ticket-splitting in the East, where those voting Republican in the Senate race chose Clinton in the presidential race, were located largely in educated suburbs of Philadelphia.  

Clinton's better overall performance than that of McGinty in transforming votes to seats is driven primarily by the East.  This is clearest at the scale of Congressional districts, where McGinty's higher vote share corresponded to a higher seat share.  In the East, on the other hand, where Clinton outpolled McGinty by 1.42 percentage points, she received a seat share that was more then 10 percentage points higher than that of McGinty.  This phenomenon persists to a lesser degree at 51 districts: both the West and East have higher mean seat shares for PRES16 than SEN16, but the difference is greater for the East (around 7 percentage points) than the West (around 4 percentage points).  It appears that Clinton's spatial pattern of support was more efficient at winning seats than McGinty's because she out-polled McGinty in marginal areas of greater Philadelphia that produced districts with small majorities for Clinton in the presidential race but small majorities for Toomey (the Republican candidate) in the Senate race.

%%%%%%%%%%%%%%

\section{Conclusion}

This chapter has focused on a single state, but we have been able to exploit useful variation of several kinds:  different vote shares and spatial patterns in different elections, different spatial scales for drawing districts, and the very different political geography of Eastern and Western Pennsylvania.  

Perhaps the most basic conclusion of this study is that because of the spatial distribution of partisanship, a neutral approach to redistricting would likely lead to the under-representation of the Democratic Party relative to its statewide strength.  In the vast majority of neutral redistricting ensembles, the Republican Party would be able to win a very comfortable majority of seats with a little less than half of the votes.  Democrats cannot expect to win a majority of seats until they win somewhere around 54 percent of the votes.  They do not benefit from a disproportionate ``winner's bonus'' until they obtain well over 56 percent of the statewide vote.  In contrast, the Republican Party can receive a massive winner's bonus even with very slightly more than 50 percent of the statewide vote.  This pattern can be seen both in Eastern and Western Pennsylvania, but it is more pronounced in the Western part of the state, where a large share of the Democrats are concentrated in Pittsburgh.

It was useful to examine a wide variety of elections not only in order to assess the implications of neutral ensembles at different statewide vote shares, but also to explore differences in the spatial support for candidates even when the overall vote shares were similar.  For instance, we discovered that in 2016, Hillary Clinton's support distribution led to a significantly better seat share than that of Katie McGinty in the Senate race, even though their statewide vote shares were quite similar.  This appears to be driven above all by Clinton's relative success in marginal suburban areas in Eastern Pennsylvania.  

This observation suggests that a state's political geography is not static, but constantly changing with time and between elections (even on the same ballot!).  Geographic realignments can and do take place.  Neutral redistricting ensembles might produce important differences in seat shares for the parties, even without large differences in statewide vote shares, if enough geographically proximate voters in marginal areas shift from one party to the other.  In many U.S. states, large swaths of suburbia have been marginally Republican for a period of time, but recent shifts in favor of the Democratic Party among educated voters in those areas---even if offset by losses in more rural areas--- could lead to changes in seat shares.  This is an important topic for further research.            

We have also explored the proposition that as the scale of districts becomes smaller, seat shares should come closer to mirroring statewide proportionality.  We explored the range from two districts to 220 districts for Pennsylvania, and found no consistent relationship between geographic scale and Republican advantage across elections.  It is entirely plausible, however, that scale effects might exist in other states over a similar range of district sizes.  In fact, we see a hint of a scale effect in Western Pennsylvania that we do not see in the East.      

We also note that the range of seat shares produced in the neutral ensembles narrows considerably as the state is divided up into more and more districts.  This leads to an interesting observation.  When the state is carved up into a relatively small number of districts, the range of outcomes is sufficiently wide that, if one draws from the most pro-Democratic tail in the distribution of plans in the ensemble, it is possible to select a plan in which 50 percent of the votes corresponds with 50 percent of the seats.  However, as the state is partitioned into smaller and smaller districts, even the most pro-Democratic plan still demonstrates substantial disadvantage for Democratic candidates.    

These findings have implications for debates about reform of redistricting in Pennsylvania and beyond. All the ensembles used here were generated by an algorithm which is independent of partisan data, and yet substantial deviations from proportionality occurred. This suggests that even a neutral process involving commissioners or demographers without access to partisan data might result in maps that lead to disproportionate results such as awarding a majority of the seats to a party that loses the statewide vote.  To be clear, our results do not show that political geography is so constraining that fair plans (as defined by measures like proportionality or a minimal efficiency gap) are impossible to draw.  Rather, some volition, based on analysis of partisan data, would be required.  

\section*{Acknowledgements}
We would like to thank Olivia Walch for the illustrations in this chapter. Thomas Weighill acknowledges the support of the NSF Convergence Accelerator Grant No. OIA-1937095.

    % uses subcaption
 

\begin{thebibliography}{xx}

\bibitem{mgggalaska2019}
Caldera, Sophia, Daryl DeFord, Moon Duchin \&\ Samuel~C. Gutekunst.
  2019.
\newblock ``Mathematics of Nested Districts: The Case of Alaska.'' {\em
  preprint} .
\newblock https://mggg.org/alaska.

\bibitem{chenrodden2013}
Chen, Jowei \&\ Jonathan Rodden. 2013.
\newblock ``Unintentional Gerrymandering: Political Geography and Electoral
  Bias in Legislatures.'' {\em Quarterly Journal of Political Science}
  8(3):239--269.

\bibitem{chenrodden2015}
Chen, Jowei \&\ Jonathan Rodden. 2015.
\newblock ``Cutting Through the Thicket: Redistricting Simulations and the
  Detection of Partisan Gerrymanders.'' {\em Election Law Journal}
  14(4):331--345.

\bibitem{pegden}
Chikina, Maria, Alan Frieze \&\ Wesley Pegden. 2017.
\newblock ``Assessing Significance in a Markov Chain Without Mixing.'' {\em
  Proceedings of the National Academy of Science} 114:2860.

\bibitem{cho_talisman}
Cho, Wendy K.~Tam \&\ Yan~Y. Liu. 2016.
\newblock ``Toward a Talismanic Redistricting Tool: A Computational Method for
  Identifying Extreme Redistricting Plans.'' {\em Election Law Journal} 15:351.
  
  \bibitem{defordrecombination} DeFord, Daryl, Moon Duchin, and Justin Solomon. "Recombination: A family of Markov chains for redistricting." arXiv preprint arXiv:1911.05725 (2019).

\bibitem{mgggvacriteria}
DeFord, Daryl \&\ Moon Duchin. 2019.
\newblock Redistricting Reform in Virginia: Districting Criteria in Context.
  Technical report MGGG.
\newblock https://mggg.org/va-criteria.pdf.

\bibitem{mgggva}
DeFord, Daryl, Moon Duchin \&\ Justin Solomon. 2018.
\newblock Comparison of districting plans for the Virginia House of Delegates.
  Technical report MGGG.
\newblock https://mggg.org/VA-report.pdf.

\bibitem{defordduchinsolomon2019}
DeFord, Daryl, Moon Duchin, and Justin Solomon. ``A Computational Approach to Measuring Vote Elasticity and Competitiveness.'' Statistics and Public Policy just-accepted (2020): 1-30. 

\bibitem{book} D. Duchin and O. Walch (eds), \emph{Political Geometry}, Birkhauser, to appear in 2021. \url{mggg.org/gerrybook}

\bibitem{duchinpa}
Duchin, Moon. 2018.
\newblock Outlier analysis for Pennsylvania congressional redistricting.
  Technical report.
\newblock https://mggg.org/uploads/md-report.pdf.

\bibitem{mgggma}
Duchin, Moon, Taissa Gladkova, Eugene Henninger-Voss, Ben Klingensmith, Heather
  Newman \&\ Hannah Wheelen. 2018.
\newblock ``Locating the representational baseline: Republicans in
  Massachusetts.'' {\em arXiv preprint arXiv:1810.09051} .

\bibitem{eubankrodden2018}
Eubank, Nicholas \&\ Jonathan Rodden. 2019.
\newblock ``Who is my Neighbor? The Spatial Efficiency of Partisanship.'' {\em
  Working Paper} .
  
  \bibitem{grofman83} Grofman, Bernard. 1983. ``Measures of bias and proportionality in seats-votes relationships''. Political Methodology,
9(3):295-327.

\bibitem{gudgintaylor79}
Gudgin, Graham \&\ P.J. Taylor. 1979.
\newblock {\em Seats, Votes, and the Spatial Organisation of Elections}.
\newblock London:  Pion Limited.

\bibitem{johnston1977}
Johnston, Ronald. 1977.
\newblock ``Spatial Structure, Plurality Systems and Electoral Bias.'' {\em The
  Canadian Geographer} 20:310--328.

\bibitem{johnston2001}
Johnston, Ronald, Charles Pattie, Dany Dorling \&\ David Rossiter.
  2001.
\newblock {\em From Votes to Seats: The Operation of the UK Electoral System
  since 1945}.
\newblock Manchester and New York:  Manchester University Press.

\bibitem{magleby}
Magleby, Daniel \&\ Daniel Mosesson. 2018.
\newblock ``A New Approach for Developing Neutral Redistricting Plans.'' {\em
  Political Analysis} 26(2):147--167.

\bibitem{amicus_math}
Mathematicians' Amicus~Brief, Rucho v. Common~Cause. 2018.
\newblock ``Amicus Brief of Mathematicians, Law Professors, and Students in
  Support of Appelleees and Affirmance.'' Amicus Brief, Supreme Court of the
  United States, Rucho et al. v. Common Cause et al.

\bibitem{mattingly}
Mattingly, Jonathan~C. \&\ Christy Vaughn. 2014.
\newblock ``Redistricting and the Will of the People.''
  https://arxiv.org/abs/1410.8796.

\bibitem{nagle2019}
Nagle, John~F. 2019.
\newblock ``What Criteria Should be Used for Redistricting Reform?'' {\em
  Election Law Journal} 18(1):63--77.

\bibitem{pegden2017}
Pegden, Wesley. 2017.
\newblock ``Pennsylvania's Congressional Districting is an Outlier: Expert
  Report.''.
\newblock Expert report submitted in Leage of Women Voters of Pennsylvania v.
  Commonwealth of Pennsylvania.

\bibitem{amicus_geographers}
Pegden, Wesley, Jonathan Rodden \&\ Samuel Wang. 2018.
\newblock ``Brief of Amici Curiae Professors Wesley Pegden, Jonathan Rodden,
  and Samuel Wang in Support of Appellees.'' Supreme Court of the United
  States.

\bibitem{rodden2019}
Rodden, Jonathan. 2019.
\newblock {\em Why Cities Lose: The Deep Roots of the Urban-Rural Divide}.
\newblock Basic Books.

\bibitem{stephanopoulos2015partisan}
Stephanopoulos, Nicholas~O \&\ Eric~M McGhee. 2015.
\newblock ``Partisan gerrymandering and the efficiency gap.'' {\em U. chi. l.
  Rev.} 82:831.

\end{thebibliography}
\end{document}